\nonstopmode
\documentclass[a4paper,UKenglish]{lipics-v2018}

\hideLIPIcs
\nolinenumbers
\usepackage{lmodern}
\usepackage{enumerate}
\usepackage{hyperref}
\usepackage{multicol}
\usepackage{amsmath}
\usepackage{proof}
\usepackage{stmaryrd}
\usepackage{latexsym}
\usepackage{amssymb}
\usepackage{amsfonts}
\usepackage{esvect}
\usepackage{alltt}
\usepackage{bussproofs}
\usepackage[all]{xy}
\usepackage{prettyref}
\usepackage{tabularx}
\usepackage{adjustbox}
\usepackage{tikz}
\usetikzlibrary{arrows}
\usepackage{xfrac}
\usepackage[symbol]{footmisc}
\usepackage{wrapfig}

\newcommand{\IF}{\mbox{ if }}
\newcommand{\AND}{\mbox{ and }}
\newcommand{\IMPLIES}{\mbox{ implies }}

\newcommand{\CD}{{\TT_{\rm CD}}}
\newcommand{\CDV}{{\TT_{\rm CDV}}}
\newcommand{\CDS}{{\TT_{\rm CDS}}}
\newcommand{\BCD}{{\TT_{\rm BCD}}}

\newcommand{\TT}{{\mathcal T}}

\newcommand{\LTT}{\lambda^{\scriptscriptstyle \rm \TT}_\cap}

\newcommand{\LCD}{\lambda^{\scriptscriptstyle \rm CD}_\cap}
\newcommand{\LCDV}{\lambda^{\scriptscriptstyle \rm CDV}_\cap}
\newcommand{\LCDS}{\lambda^{\scriptscriptstyle \rm CDS}_\cap}
\newcommand{\LBCD}{\lambda^{\scriptscriptstyle \rm BCD}_\cap}

\newcommand{\DD}[2]{\Delta^{\scriptscriptstyle \rm #1}_{\scriptscriptstyle #2}}

\newcommand{\DCD}{\Delta^{\scriptscriptstyle \rm CD}_\R}
\newcommand{\DCDV}{\Delta^{\scriptscriptstyle \rm CDS}_\R}
\newcommand{\DCDS}{\Delta^{\scriptscriptstyle \rm CDV}_\R}
\newcommand{\DBCD}{\Delta^{\scriptscriptstyle \rm BCD}_\R}

\newcommand {\ala}{{\textit{\`a la}}}
\newcommand {\ie}{{\textit{i.e.}}}
\newcommand {\eg}{{\textit{e.g.}}}
\newcommand {\viz}{{\textit{viz.}}}
\newcommand {\vs}{{\textit{vs.}}}
\newcommand {\etal}{{\textit{et al.}}}
\newcommand {\adhoc}{{\textit{ad hoc}}}

\newcommand{\Adef}{\mathbb{A}}
\newcommand{\A}{\mathbb{A_{\infty}}}
\newcommand{\Aom}{\mathbb{A^{\om}_{\infty}}}

\newcommand{\red}{\longrightarrow}
\newcommand{\multired}{{\red\!\!\!\!\!\!\!\!\!\red}}

\newcommand{\taka}{\Longrightarrow}

\newcommand{\at}{\,}
\newcommand{\eqdef}{\stackrel{\mathit{def}}{=}}
\newcommand{\of}{{:}}
\renewcommand{\leq}{\leqslant}
\newcommand{\R}{\mathcal{R}}

\newcommand{\subst}[3]{{#1}[{#3}/{#2}]}

\newcommand{\leqmin}{\leq_{\text{min}}}
\newcommand{\essence}[1]{\mathopen \wr\,#1\,\mathclose \wr}

\newcommand{\trans}[1]{\left\|{#1}\right\|}
\newcommand{\map}[1]{| #1 |}
\newcommand{\pair}[2]{\ensuremath{( {#1}\:,{#2} ) }}

\newcommand{\spair}[2]{\ensuremath{\langle {#1}\:,{#2} \rangle }}

\newcommand{\prl}{{pr}_{\!1}\,}
\newcommand{\prr}{{pr}_{\!2}\,}
\newcommand{\pri}{{pr}_{\!i}\,}

\newcommand{\s}{\ensuremath{\sigma}}
\renewcommand{\t}{\ensuremath{\tau}}
\renewcommand{\r}{\ensuremath{\rho}}
\renewcommand{\omega}{\ensuremath{{\tt U}}}
\newcommand{\om}{\ensuremath{\omega}}

\newcommand{\D}{\ensuremath{\Delta}}

\title{The $\D$-calculus: Syntax and Types}

\titlerunning{The $\D$-calculus: Syntax and Types}

\author{{Luigi Liquori \qquad  Claude Stolze} \footnote{Work supported by the COST Action CA15123 EUTYPES “The European research network on types for programming and verification”.}} 
{Universit\'{e} C\^{o}te d'Azur, Inria, France}
{[Luigi.Liquori,Claude.Stolze]@inria.fr}{}{}

\authorrunning{Luigi Liquori and  Claude Stolze}

\Copyright{Luigi Liquori and Claude Stolze}


\acknowledgements{We are grateful to  Benjamin Pierce and Furio Honsell for the useful comments and remarks.
}

\begin{document}

\maketitle

\begin{abstract}
  We present the $\D$-calculus, an explicitly typed $\lambda$-calculus
  with strong pairs, projections and explicit type coercions. The
  calculus can be parametrized with different intersection type theories
  $\TT$, \eg\ the Coppo-Dezani, the Coppo-Dezani-Sall\'e, the
  Coppo-Dezani-Venneri and the Barendregt-Coppo-Dezani ones, producing a
  family of $\D$-calculi with related intersection typed systems. We
  prove the main properties like Church-Rosser, unicity of type, subject
  reduction, strong normalization, decidability of type checking and
  type reconstruction. We state the relationship between the
  intersection type assignment systems \ala\ Curry and the corresponding
  intersection typed systems \ala\ Church by means of an essence function
  translating an explicitly typed $\D$-term into a pure $\lambda$-term
  one. We finally translate a $\D$-term with type coercions into an
  equivalent one without them; the translation is proved to be coherent
  because its essence is the identity. The generic $\D$-calculus can be
  parametrized to take into account other intersection type theories as
  the ones in the Barendregt \etal\ book.
\end{abstract}

\maketitle


\vspace{-3mm} \section{Introduction} \vspace{-3mm}

Intersection type theories $\TT$ were first
introduced as a form of \adhoc\ polymorphism in (pure)
$\lambda$-calculi \ala\ Curry. The paper by Barendregt, Coppo, and
Dezani \cite{BCD} is a classic reference, while \cite{bar2013} is a
definitive reference.

Intersection type assignment systems $\LTT$ have been well-known in
the literature for almost 40 years for many reasons: among them,
characterization of strongly normalizing $\lambda$-terms
\cite{bar2013}, $\lambda$-models \cite{ABD06}, automatic type
inference \cite{KW04}, type inhabitation \cite{urzy99,refurzy12}, type
unification \cite{DMR17}. As intersection had its classical
development for type assignment systems, many papers tried to find an
explicitly typed $\lambda$-calculus \ala\ Church corresponding to the
original intersection type assignment systems \ala\ Curry. The
programming language Forsythe, by Reynolds \cite{reyn88}, is probably
the first reference, while Pierce's Ph.D. thesis \cite{pier91b}
combines also unions, intersections and bounded polymorphism. In
\cite{wells2} intersection types were used as a foundation for typed
intermediate languages for optimizing compilers for higher-order
polymorphic programming languages; implementations of typed
programming language featuring intersection (and union) types can be
found in SML-CIDRE \cite{davies-phd} and in StardustML
\cite{dunfield07,Dunfield14}.

Annotating pure $\lambda$-terms with intersection types is not simple:
a classical example is the difficulty to decorate the bound variable
of the explicitly typed polymorphic identity $\lambda x\of?.x$ such
that the type of the identity is $(\s\to\s) \cap (\t\to\t)$: previous
attempts showed that the full power of the intersection type
discipline can be easily lost.

In this paper, we define and prove the main properties of the
$\D$-calculus, a generic intersection typed system for an explicitly
typed $\lambda$-calculus \ala\ Church enriched with strong pairs,
denoted by $\spair{\D_1}{\D_2}$, projections, denoted by $\pri \D$,
and type coercions, denoted by $\D^\s$.

A \emph{strong pair} $\spair{\D_1}{\D_2}$ is a special kind of
cartesian product such that the two parts of a pair satisfies a given
property $\R$ on their ``essence'', that is
$\essence{\D_1} \mathrel{\R} \essence{\D_2}$.

An \emph{essence} $\essence{\D}$ of a $\D$-term is a pure
$\lambda$-term obtained by erasing type decorations, projections and
choosing one of the two elements inside a strong pair. As examples,
{ \begin{eqnarray*} \essence{\spair{\lambda x\of\s\cap\t.\prr x}{\lambda
      x\of\s\cap\t.\prl x}} & = & \lambda x.x\\ \essence{\lambda x \of
    (\s \to \t) \cap \s. (\prl x) (\prr x)} & = & \lambda x.x\at x\\
  \essence{ \lambda x\of \s \cap (\t\cap \r).\spair{\spair{\prl x
      }{\prr \prl x}}{\prr \prr x}} & = & \lambda x.x
\end{eqnarray*}}
 and so on. Therefore, the essence of a $\D$-term is
its untyped skeleton: a strong pair $\spair{\D_1}{\D_2}$ can be
typechecked if and only if $\essence{\D_1} \mathrel{\R}
\essence{\D_2}$ is verified, otherwise the strong pair will be
ill-typed. The essence also gives the exact mapping between a term and
its typing \ala\ Church and its corresponding term and type assignment
\ala\ Curry.  Changing the parameters $\TT$ and $\R$ results in
defining a totally different intersection typed system.  For the
purpose of this paper, we study the four well-known intersection type
theories $\TT$, namely Coppo-Dezani $\CD$ \cite{CD},
Coppo-Dezani-Sall\'e $\CDS$ \cite{CDS}, Coppo-Dezani-Venneri $\CDV$
\cite{CDV} and Barendregt-Coppo-Dezani $\BCD$ \cite{BCD}. We will
inspect the above type theories using three equivalence relations $\R$
on pure $\lambda$-terms, namely $\equiv, =_\beta$ and $=_{\beta\eta}$.

The combination of the above $\TT$ and $\R$ allows to define ten
meaningful typed systems for the $\Delta$-calculus that can be
pictorially displayed in a ``$\D$-chair'' (see Definition
\ref{chair}). Following the same style as in the Barendrengt \etal\
book \cite{bar2013}, the edges in the chair represent an inclusion
relation over the set of derivable judgments.

A type coercion $\D^\t$ is a term of type $\t$ whose type-decoration
denotes an application of a subsumption rule to the term $\D$ of type
$\s$ such that $\s \leq_\TT \t$: if we omit type coercions, then we
lose the uniqueness of type property.

Section \ref{exa} shows a number of typable examples in the
systems presented in the $\D$-chair: each example is provided with a
corresponding type assignment derivation of its essence. Some
historical examples of Pottinger \cite{Pottinger-80}, Hindley
\cite{Hindley82} and Ben-Yelles \cite{Ben-Yelles} are essentially
re-decorated and inhabited (when possible) in the $\D$-calculus. The
aims of this section is both to make the reader comfortable with the
different intersection typed systems, and to give a first intuition of
the correspondence between Church-style and Curry-style calculi.

Section \ref{meta} proves the metatheory for all the
systems in the $\D$-chair: Church-Rosser, unicity of type, subject
reduction, strong normalization, decidability of type checking and
type reconstruction and studies the relations between
intersection type assignment systems \ala\ Curry and the corresponding
intersection typed systems \ala\ Church. Notions of soundness,
completeness and isomorphism will relate type assignment and typed
systems. We also show how to get rid of type coercions $\D^\t$
defining a translation function, denoted by $\trans{\_}$, inspired by
the one of Tannen \etal\ \cite{TannenCGS91}: the intuition of the
translation is that if $\D$ has type $\s$ and $\s \leq_\TT \t$, then
$\trans{\s \leq_\TT \t}$ is a $\D$-term of type $\s \to \t$,
$(\trans{\s \leq_\TT \t} \at \trans{\D})$ has type $\t$ and
$\essence{\trans{\s \leq_\TT \t}}$ is the identity $\lambda x.x$.

\vspace{-2mm}\subsection{$\lambda$-calculi with intersection types \ala\
Church} \vspace{-2mm}
Several calculi \ala\ Church appeared in the literature: they 
capture the power of intersection types; we briefly review
them. 

The Forsythe programming language by Reynolds \cite{reyn88}
annotates a $\lambda$-abstraction with types as in
$\lambda x\of\s_1 {\mid} {\cdots} {\mid} \s_n.M$. However, we cannot
type a typed term, whose type erasure is the combinator
${\sf K} \equiv \lambda x.\lambda y.x$, with the type
$(\s\to\s\to\s) \cap (\t\to\t\to\t)$.

Pierce \cite{pier91a} improves Forsythe by using a {\bf for}
construct to build \adhoc\ polymorphic typing, as in
${\bf for}\,{\alpha \in\{\s,\t\}.\lambda x\of\alpha,\lambda
  y\of\alpha.x}$. However, we cannot type a typed term, whose type
erasure is $\lambda x.\lambda y.\lambda z. \pair{x\at y}{x\at z}$,
with the type
\\$ ((\s\to\r)\cap(\t\to\r') \to \s\to\t\to\r\times\r') \cap
  ((\s\to\s)\cap(\s\to\s) \to \s\to\s\to\s\times\s).
$

Freeman and Pfenning \cite{PF91} introduced refinement types,
that is types that allow \emph{ad hoc} polymorphism for ML
constructors. Intuitively, refinement types can be seen as subtypes of
a standard type: the user first defines a type and then the refinement
types of this type. The main motivation for these refinement types is
to allow non-exhaustive pattern matching, which becomes exhaustive for
a given refinement of the type of the argument. As an example, we can
define a type {\small\verb|boolexp|} for boolean expressions, with
constructors {\small\verb|True|}, {\small\verb|And|},
{\small\verb|Not|} and {\small\verb|Var|}, and a refinement type
{\small\verb|ground|} for boolean expressions without variables, with
the same constructors except {\small\verb|Var|}: then, the constructor
{\small \verb|True|} has type {\small
  $\verb|boolexp| \cap \verb|ground|$}, the constructor {\small
  \verb|And|} has type {\small
  $(\verb|boolexp| * \verb|boolexp| \to \verb|boolexp|) \cap
  (\verb|ground| * \verb|ground| \to \verb|ground|)$} and so
on. However, intersection is meaningful only when using constructors.

Wells \etal\ \cite{wells2} introduced $\lambda^{\rm CIL}$, a typed intermediate
$\lambda$-calculus for optimizing compilers for higher-order
programming languages. The calculus features intersection, union and
flow types, the latter being useful to optimize data
representation. $\lambda^{\rm CIL}$ can faithfully encode an
intersection type assignment derivation by introducing the concept of
virtual tuple, \ie\ a special kind of pair whose type erasure leads to
exactly the same untyped $\lambda$-term. A parallel context and
parallel substitution, similar to the notion of \cite{LR05,LR07}, is
defined to reduce expressions in parallel inside a virtual
tuple. Subtyping is defined only on flow types and not on intersection
types: this system can encode the $\LCD$ type assignment system.


Wells and Haak \cite{wells3} introduced $\lambda^{\rm B}$, a
more compact typed calculus encoding of $\lambda^{\rm CIL}$: in fact,
by comparing Fig. 1 and Fig. 2 of \cite{wells3} we can see that the
set of typable terms with intersection types of $\lambda^{\rm CIL}$
and $\lambda^{\rm B}$ are the same. In that paper, virtual tuples are
removed by introducing branching terms, typable with branching types,
the latter representing intersection type schemes. Two operations on
types and terms are defined, namely {\sf expand}, expanding the
branching shape of type annotations when a term is substituted into a
new context, and {\sf select}, to choose the correct branch in terms
and types. As there are no virtual tuples, reductions do not need to
be done in parallel.  As in \cite{wells2}, the $\LCD$ type assignment system
can be encoded.

Frisch \etal\ \cite{Frisch08} designed a typed system with
intersection, union, negation and recursive types. 
The authors inherit the usual problem of having a domain space $\mathcal D$ that
contains all the terms and, at the same time, all the functions from
$\mathcal D$ to $\mathcal D$. 
They prevent this by having an auxiliary domain space which is the disjoint
union of $\mathcal D^2$ and $\mathcal P(\mathcal D^2 )$. 
The authors interpret types as sets in a well-suited model where the 
set-inspired type constructs are interpreted as the corresponding to 
set-theoretical constructs. Moreover, the model manages higher-order 
functions in an  elegant way.
The subtyping relation is defined as a relation on the set-theoretical interpretation
$\llbracket \_ \rrbracket$ of the types. For instance, the problem
$\s\cap\t\leq\s$ will be interpreted as
$\llbracket \s \rrbracket \cap \llbracket \t \rrbracket \subseteq
\llbracket \s \rrbracket$, where $\cap$ becomes the set intersection
operator, \mbox{and the decision program actually decides whether
$(\llbracket \s \rrbracket \cap \llbracket \t \rrbracket) \cap
\overline{\llbracket \s \rrbracket}$ is the empty set.}

Bono \etal\ \cite{BVB} introduced a relevant and strict
parallel term constructor to build inhabitants of intersections and a
simple call-by-value parallel reduction strategy. An infinite number
of constants $c^{\s\Rightarrow\t}$ is applied to typed variables
$x^\s$ such that $c^{\s\Rightarrow\t} \at x^\s$ is upcasted to type
$\t$.  It also uses a local renaming typing rule, which
changes type decoration in $\lambda$-abstractions, as well as
coercions. Term synchronicity in the tuples is guaranteed by the
typing rules. The calculus uses van Bakel's strict version
\cite{bakel2004} of the $\CD$ intersection type theory.

\vspace{-2mm}\subsection{Logics for intersection types} \vspace{-2mm}
Proof-functional (or strong) logical connectives, introduced by Pottinger
\cite{Pottinger-80}, take into account the shape of logical proofs,
thus allowing for polymorphic features of proofs to be made explicit
in formul\ae. This differs from classical or intuitionistic
connectives where the meaning of a compound formula is only dependent
on the truth value or the provability of its subformul\ae.

Pottinger was the first to consider the intersection $\cap$ as a
proof-functional connective. He contrasted it to the intuitionistic
connective $\wedge$ as follows: \emph{``The intuitive meaning of
  $\cap$ can be explained by saying that to assert $A \cap B$ is to
  assert that one has a reason for asserting $A$ which is also a
  reason for asserting $B$, while to assert $A \wedge B$ is to assert
  that one has a pair of reasons, the first of which is a reason for
  asserting $A$ and the second of which is a reason for asserting
  $B$''}.

A simple example of a logical theorem involving intuitionistic
conjunction which does not hold for proof-functional conjunction is
$(A\supset A) \wedge (A \supset B \supset A)$. Otherwise there would
exist a term which behaves both as $\sf I$ and as $\sf K$. Later,
Lopez-Escobar \cite{Lopez-Escobar85} and Mints \cite{Mints89}
investigated extensively logics featuring both proof-functional and
intuitionistic connectives especially in the context of realizability
interpretations.

It is not immediate to extend the judgments-as-types Curry-Howard
paradigm to logics supporting proof-functional connectives. These
connectives need to compare the shapes of derivations and do not just
take into account their provability, \ie\ the inhabitation of the
corresponding type.

There are many proposals to find a suitable logics to fit intersection
types; among them we cite
\cite{venneri94,roncrove01,miquel01,CLV,BVB,pimronrov12}, and previous
papers by the authors \cite{APLAS16,TTCS17,LFMTP17}. 


\vspace{-2mm}\subsection{Raising the $\D$-calculus to a $\D$-framework.}  \vspace{-2mm}
Our goal is  to build a prototype of a theorem prover
based on the $\D$-calculus and proof-functional logic. 
Recently \cite{FSTTCS18}, we have extended a subset of the generic $\D$-calculus 
with other proof-functional operators like 
union types, relevant arrow types, together with  dependent types as in the 
Edinburgh Logical Framework \cite{LF}: a preliminary implementation of a type 
checker appeared in \cite{LFMTP17} by the authors. In a nutshell:

\emph{Strong disjunction} is a proof-functional connective that can be
interpreted as the union type $\cup$ \cite{APLAS16,LFMTP17}: it
contrasts with the intuitionistic connective $\vee$. As Pottinger did
for intersection, we could say that asserting $(A \cup B) \supset C$
is to assert that one has a reason for $(A \cup B) \supset C$, which
is also a reason to assert $A \supset C$ and $B \supset C$.  A simple
example of a logical theorem involving intuitionistic disjunction
which does not hold for strong disjunction is
$((A \supset B) \cup B) \supset A \supset B$. \mbox{Otherwise there would
exist a term which behaves both as $\sf I$ and as $\sf K$.}

\emph{Strong (relevant) implication} is yet another proof-functional
connective that was interpreted in \cite{BM94} as a relevant arrow
type $\to_r$. As explained in \cite{BM94}, it can be viewed as a
special case of implication whose related function space is the
simplest one, namely the one containing only the identity
function. \mbox{Because the operators $\supset$ and $\to_r$ differ,
$A \to_r B \to_r A$ is not derivable.}

\emph{Dependent types}, as introduced in the Edinburgh Logical
Framework \cite{LF} by Harper \etal, allows considering proofs as
first-class citizens albeit differently with respect to
proof-functional logics. The interaction of both dependent and
proof-functional operators is intriguing: the former mentions proofs
explicitly, while the latter mentions proofs implicitly. Their
combination therefore opens up new possibilities of formal reasoning
on proof-theoretic semantics.



\vspace{-3mm} \section{Syntax, Reduction and Types}\label{sys} \vspace{-3mm}

\begin{figure*}[t]
  $
    \begin{array}[c]{llll} \multicolumn{4}{l}{{\bf
      Minimal~type~theory} \leqmin}\\[1mm]
      ({\rm refl}) & \s \leq \s
      &
        ({\rm incl}) & \s \cap \t \leq \s,
                       \s \cap \t \leq \t
      \\[1mm]
      ({\rm glb}) & \r \leq \s, \r \leq \t \Rightarrow
                    \r \leq \s \cap \t
      &
        ({\rm trans}) & \s \leq \t, \t \leq \r \Rightarrow
                        \s \leq \r
      \\[1mm]
      \multicolumn{4}{l}{\bf Axiom~schemes}\\[1mm]
      (\om_{top}) & \s \leq \omega
      &
        (\om_\rightarrow) & \omega \leq \s \to \omega
      \\[1mm]
      ({\rightarrow}{\cap}) & (\s \to \t) \cap (\s \to \r) \leq
                              \s \to (\t \cap \r)
      \\[1mm]
      \multicolumn{4}{l}{\bf Rule~scheme}\\[1mm]
      (\rightarrow) & \s_2 \leq \s_1, \t_1 \leq \t_2 \Rightarrow
                      \s_1 \to \t_1 \leq \s_2 \to \t_2
    \end{array}
  $\vspace{-2mm}
  \caption{Minimal type theory $\leqmin$, axioms and rule schemes
    (see Fig. 13.2 and 13.3 of \cite{bar2013})}
  \label{subtype}
\end{figure*}

\begin{figure*}[t]
  $
    \begin{array}[c]{rr}
      \infer[(ax)]
      {B \vdash^\TT_{\cap} x : \s}
      {x \of \s \in B}
      &
        \infer[({\rightarrow}I)]
        {B \vdash^\TT_{\cap} \lambda x.M : \s \to \t}
        {B,x\of\s \vdash^\TT_{\cap} M:\t}
      \\[2mm]
      \infer[(\cap I)]
      {B \vdash^\TT_{\cap} M : \s \cap \t}
      {B \vdash^\TT_{\cap} M : \s
      &
        B \vdash^\TT_{\cap} M : \t}
      &
        \infer[({\rightarrow}E)] {B \vdash^\TT_{\cap} M \at N : \t}
        {B \vdash^\TT_{\cap} M : \s \to \t & B \vdash^\TT_{\cap} N : \s}
      \\[2mm]
      \infer[(\cap E_1)]
      {B \vdash^\TT_{\cap} M : \s}
      {B \vdash^\TT_{\cap} M : \s \cap \t}
      &
        \infer[(\cap E_2)]
        {B \vdash^\TT_{\cap} M : \t}
        {B \vdash^\TT_{\cap} M : \s \cap \t}
      \\[2mm]
      \infer[(top)]
      {B \vdash^\TT_{\cap} M : \omega}
      {\om \in \Adef}
      &
        \infer[(\leq_\TT)]
        {B \vdash^\TT_{\cap} M : \t}
        {B \vdash^\TT_{\cap} M : \s & \s \leq_\TT \t}
    \end{array}
  $\vspace{-2mm}
  \caption{Generic intersection type assignment system $\LTT$
    (see Figure 13.8 of \cite{bar2013})}
  \label{curry}\vspace{-4mm}
\end{figure*}

\begin{definition}[Type atoms, type syntax, type theories and type
  assignment systems]\label{tas} 
  We briefly review some basic definition from Subsection 13.1 of
  \cite{bar2013}, in order to define type assignment systems. The set
  of atoms, intersection types, intersection type theories and
  intersection type assignment systems are defined as follows:
  \begin{enumerate}
  \item {\bf (Atoms).} Let $\Adef$ denote a set of symbols which we
    will call type atoms, and let $\omega$ be a special type atom
    denoting the universal type. In particular, we will use
    $\A = \{{\tt a}_i \mid i\in\mathbb N\}$ with ${\tt a}_i$ being
    different from $\om$ and $\Aom = \A \cup \{\omega\}$.
  \item {\bf (Syntax).} The syntax of intersection types, parametrized
    by $\Adef$, is:
    $\s ::= \Adef \mid \s \to \s \mid \s \cap \s$.
  \item {\bf (Intersection type theories $\TT$).} \label{tt} An
    intersection type theory $\TT$ is a set of sentences of the form
    $\s\leq\t$ satisfying at least the axioms and rules of the minimal
    type theory $\leqmin$ defined in Figure \ref{subtype}. The type
    theories $\CD,\CDV,\CDS$, and $\BCD$ are the smallest type theories over $\Adef$
    satisfying the axioms and rules given in Figure \ref{subtype2}. We
    write $\TT_1 \sqsubseteq \TT_2$ if, for all $\s,\t$ such that
    $\s \leq_{\TT_1} \t$, we have that $\s \leq_{\TT_2} \t$. In
    particular $\CD \sqsubseteq \CDV \sqsubseteq \BCD$ and
    $\CD \sqsubseteq \CDS \sqsubseteq \BCD$. We will sometime note,
    for instance, {\small $\rm BCD$} instead of $\BCD$.
  \item {\bf (Intersection type assignment systems $\LTT$).} We define
    in Figure \ref{curry}\footnote[4]{Although rules $(\cap E_i)$ are
      derivable with $\leqmin$, we add them for clarity.} an infinite
    collection of type assignment systems parametrized by a set of
    atoms $\Adef$ and a type theory $\TT$. We name four particular
    type assignment systems in the table below, which is an excerpt
    from Figure 13.4 of \cite{bar2013}. 
    $B \vdash^\TT_\cap M : \s$ denotes a derivable type
    assignment judgment \mbox{in the type assignment system $\LTT$. Type
    checking is not decidable for $\LCD$, $\LCDV$, $\LCDS$, and
    $\LBCD$.}

    \begin{figure*}[t]
      \[
        \begin{array}{| l | l | l | l | l |}
          \hline
          \LTT & \TT & \Adef
          & \leqmin {\rm plus} & {\rm ref.}
          \\\hline\hline
          \LCD & \CD & \A
          & - & \mbox{\cite{CD}}
          \\\hline
          \LCDS & \CDS & \Aom
          & (\om_{top}) & \mbox{\cite{CDS}}
          \\\hline
          \LCDV & \CDV & \A
          & (\to),({\to}\cap) & \mbox{\cite{CDV}}
          \\\hline
          \LBCD & \BCD & \Aom
          & (\to),({\to}\cap),(\om_{top}),(\omega{\to}) & \mbox{\cite{BCD}}
          \\ \hline
        \end{array}
      \]
      \caption{Type theories $\LCD$, $\LCDS$, $\LCDV$, and
        $\LBCD$. The ref. column refers to the original article these
        theories come from.}
      \label{subtype2}
    \end{figure*}
  \end{enumerate}
\end{definition}

\vspace{-3mm} \subsection{The $\D$-calculi} \vspace{-2mm}
Intersection type assignment systems and $\D$-calculi have in common
their type syntax and intersection type theories. The generic syntax
of the $\D$-calculus is defined as follows.
\begin{definition}[Generic $\D$-calculus syntax]\label{SYNTAX}
  \[
    \begin{array}{rcl}
      \D & ::=
      &
        u_\D \mid x \mid
        \lambda x \of \s. \D \mid
        \D \at \D \mid
        \spair{\D}{\D} \mid
        \pri \D
        \mid
        \D^\s
        \qquad\qquad i \in \{1,2\}
    \end{array}
  \]
\end{definition}
$u_\D$ denotes an infinite set of constants, indexed with a particular
untyped $\D$-term. $\D^\s$ denotes an explicit coercion of $\D$ to
type $\s$. The expression $\spair{\D}{\D}$ denotes a pair that,
following the Lopez-Escobar jargon \cite{Lopez-Escobar85}, we call
``strong pair'' with respective projections $\prl$ and $\prr$.
The essence function $\essence{\_}$ is an erasing function mapping
typed $\D$-terms into pure $\lambda$-terms. It is defined as follows.

\begin{definition}[Essence function]\label{essence}  
  \[
    \begin{array}{rcl@{\qquad\qquad}rcl@{\qquad}rcl}
      \essence{x} & \eqdef & x
      &
        \essence{\D^\s}& \eqdef & \essence{\D}
        &
      \essence{u_\D} & \eqdef & \essence{\D}
      \\[2mm]
      \essence{\lambda x \of \s. \D} & \eqdef & \lambda x. \essence{\D}
      &
        \essence{\D_1 \at \D_2} & \eqdef & \essence{\D_1} \at \essence{\D_2}
      \\[2mm]
      \essence{\spair{\D_1}{\D_2}} & \eqdef & \essence{\D_1}
      &
        \essence{\pri \D} & \eqdef & \essence{\D} \qquad i\in\{1,2\}
    \end{array}
  \]
\end{definition}
One could argue that the choice of
$\essence{\spair{\D_1}{\D_2}} \eqdef \essence{\D_1}$ is arbitrary and
could have been replaced with
$\essence{\spair{\D_1}{\D_2}} \eqdef \essence{\D_2}$. However, the
typing rules will ensure that, if $\spair{\D_1}{\D_2}$ is typable,
then, for some suitable equivalence relation $\R$, we have that 
$\essence{\D_1} \mathrel{\R} \essence{\D_2}$.  Thus, strong pairs
can be viewed as constrained cartesian products.
The generic reduction semantics reduces terms of the $\D$-calculus as
follows.
\begin{definition}[Generic reduction semantics]\label{red} 
  Syntactical equality is denoted by $\equiv$.
  \begin{enumerate}
  \item {\bf (Substitution)} Substitution on $\D$-terms is defined as
    usual, with the additional rules:
    \[
        \subst{u_{\D_1}}{x}{\D_2}  \eqdef  u_{(\subst{\D_1}{x}{\D_2})}
       \qquad  \mbox{and} \qquad 
        \subst{\D_1^\s}{x}{\D_2}  \eqdef  (\subst{\D_1}{x}{\D_2})^\s
    \]
  \item {\bf (One-step reduction).}
    We define three notions of reduction:
    \[
      \begin{array}{rcl@{\qquad}l}
        (\lambda x \of \s. \D_1) \at \D_2 & \red & \subst{\D_1}{x}{\D_2} & (\beta)
        \\[2mm]
        \pri \spair{\D_1}{\D_2} & \red & \D_i \quad i \in\{1,2\} & (\pri)
        \\[2mm]
        \lambda x \of \s. \D \at x & \red & \D \quad x \not \in {\rm FV}(\D) & (\eta)
      \end{array}
    \]
    Observe that $(\lambda x\of\s.\D _1)^\s \at \D _2$
    is not a redex, because the $\lambda$-abstraction is coerced. 
    The contextual closure is defined as usual except for 
    reductions inside the index of $u_\D$ that are forbidden (even though
    substitutions are propagated). We write
    $\red_{\beta\pri}$ for the contextual closure of the $(\beta)$ and
    $(\pri)$ notions of reduction, $\red_\eta$ for the contextual
    closure of $(\eta)$.  We also define a synchronous
    contextual closure, which is like the usual contextual closure
    except for the strong pairs, as defined in point
    $(3)$. Synchronous contextual closure of the notions of reduction
    generates the reduction relations $\red^\parallel_{\beta\pri}$ and 
    $\red^\parallel_\eta$.
  \item {\bf (Synchronous closure of $\red^\parallel$).} Synchronous
    closure 
    is defined on the strong pairs with the following constraint:\\
    $
      \infer[(Clos^{\parallel})] {\spair{\D_1}{\D_2} \red^{\parallel}
        \spair{\D'_1}{\D'_2}} {\D_1\red^\parallel \D'_1 & \D_2
        \red^\parallel \D'_2 & \essence{\D'_1} \equiv
        \essence{\!\D'_2}}
    $
    \\
    Note that we reduce in the two components of the strong pair;

\item {\bf (Multistep reduction)}. We write $\multired_{\beta\pri}$ (resp. $\multired_{\beta\pri}^\parallel$) as the reflexive and transitive closure of $\red_{\beta\pri}$ (resp. $\red_{\beta\pri}^\parallel$);   
    
\item {\bf (Congruence)}. We write $=_{\beta\pri}$ as the symmetric, reflexive, transitive closure of $\multired_{\beta\pri}$. 
    
  \end{enumerate}
We mostly consider $\beta\pri$-reductions, thus to ease the notation we omit the subscript in $\beta\pri$-reductions. 
\end{definition}



The next definition introduces a notion of synchronization inside
strong pairs.
\begin{definition}[Synchronization\label{def-sync}] 
  A $\D$-term is synchronous if and only if, for all its subterms of
  the shape $\spair{\D_1}{\D_2}$, we have that
  $\essence{\D_1} \equiv \essence{\D_2}$.
\end{definition}
\begin{figure*}[t]
  \[
    \begin{array}{rrr}
      \infer[(top)]
      {B \vdash^\TT_{\R} u_\D : \om}
      {\om \in \Adef}
      \hfill
      \infer[(ax)]
      {B \vdash^\TT_{\R} x : \s}
      {x \of \s \in B}
      &
        \infer[({\to}I)]
        {B \vdash^\TT_{\R} \lambda x\of \s.\D : \s \to \t}
        {B,x\of\s \vdash^\TT_{\R} \D:\t}
      \\[2mm]
      \infer[(\cap I)]
      {B \vdash^\TT_{\R} \spair{\D_1}{\D_2} : \s \cap \t}
      {B \vdash^\TT_{\R} \D_1 : \s
      &
        B \vdash^\TT_{\R} \D_2 : \t
      &
        \mathop{\essence{\D_1}} \mathrel{\R} \mathop{\essence{\D_2}}}
      &
        \infer[({\to}E)]
        {B \vdash^\TT_{\R} \D_1 \at \D_2 : \t}
        {B \vdash^\TT_{\R} \D_1: \s \to \t & B \vdash^\TT_{\R} \D_2: \s}
      \\[2mm]
      \infer[(\cap E_1)]
      {B \vdash^\TT_{\R} \prl{\D} : \s}
      {B \vdash^\TT_{\R} \D : \s \cap \t}
      \hfill
      \infer[(\cap E_2)]
      {B \vdash^\TT_{\R} \prr{\D} : \t}
      {B \vdash^\TT_{\R} \D : \s \cap \t}
      &
        \infer[(\leq_\TT)]
        {B \vdash^\TT_{\R} \D^\t : \t}
        {B \vdash^\TT_{\R} \D : \s & \s \leq_\TT \t}
    \end{array}
  \]
  \caption{Generic intersection typed system $\DD{\TT}{\R}$}
  \label{CHURCH}
\end{figure*}
\mbox{It is easy to verify that $\red^\parallel$ preserves
synchronization, while it is not the case for $\red$.
The} next definition introduces a generic intersection typed system for
the $\D $-calculus that is \mbox{parametrizable by suitable equivalence
relations on pure $\lambda$-terms $\R$ and type theories $\TT$ as
follows.}

\begin{definition}[Generic intersection typed system]\label{church} 
  The generic intersection typed system is defined in Figure
  \ref{CHURCH}.  We denote by $\DD{\TT}{\R}$ a particular typed system
  with the type theory $\TT$ and under an equivalence relation $\R$
  and by $B \vdash^\TT_\R \D : \s$ a corresponding typing judgment.
\end{definition}
The typing rules are intuitive for a calculus \ala\ Church
except rules $(\cap I)$, $(top)$ and $(\leq_\TT)$.

The typing rule for a strong pair $(\cap I)$ is similar to the typing
rule for a cartesian product, except for the side-condition
$\essence{\D_1} \mathrel{\R} \essence{\D_2}$, forcing the two parts of
the strong pair to have essences compatible under $\R$, thus making a
strong pair a special case of a cartesian pair. For instance,
$\spair{\lambda x\of\s.\lambda y\of \t.x}{\lambda x\of\s.x}$ is not 
typable in $\DD{\TT}{\equiv}$;  $\spair{(\lambda x\of\s.x) \at y}{y}$ is
not typable in $\DD{\TT}{\equiv}$ but it is in $\DD{\TT}{=_\beta}$;
$\spair{x}{\lambda y\of \s. ((\lambda z\of\t.z) \at x) \at y}$ is not
typable in  $\DD{\TT}{\equiv}$ nor $\DD{\TT}{=_\beta}$ but it is in
$\DD{\TT}{=_{\beta\eta}}$. 
In the typing rule $(top)$, the subscript $\D$ in $u_\D$ is not
necessarily typable so $\essence{u_\D}$ can easily be any arbitrary
$\lambda$-term.
%
%
The typing rule $(\leq_\TT)$ allows to change the type of a $\D$-term
from $\s$ to $\t$ if $\s \leq_\TT \t$: the term in the conclusion must
record this change with an explicit type coercion $\_^\t$, producing
the new term $\D^\t$: explicit type coercions are important to keep
the unicity of typing derivations.

The next definition introduces a partial order over equivalence
relations on pure $\lambda$-terms and an inclusion over typed systems
as follows.

\begin{definition}[$\R$ and $\sqsubseteq$] 
  \begin{enumerate}

  \item Let $\R \in \{\equiv,=_\beta,=_{\beta\eta}\}$.
    $\R_1 \sqsubseteq \R_2$ if, for all pure $\lambda$-terms $M,N$
    such that $M \mathrel{\R_1} N$, we have that $M \mathrel{\R_2} N$;
  \item if $B \vdash^{\TT_1}_{\R_1} \D:\s$ and
    $\DD{\TT_1}{\R_1} \sqsubseteq \DD{\TT_2}{\R_2}$, then
    $B \vdash^{\TT_2}_{\R_2} \D:\s$.
  \end{enumerate}
\end{definition}

\begin{lemma}\label{include}
  \begin{enumerate}
  \item $\DCD \sqsubseteq \DCDV \sqsubseteq \DBCD$ and
    $\DCD \sqsubseteq \DCDS \sqsubseteq \DBCD$;
  \item $\DD{\TT_1}{\R_1} \sqsubseteq \DD{\TT_2}{\R_2}$ if
    $\TT_1 \sqsubseteq \TT_2$ and $\R_1 \sqsubseteq \R_2$.
  \end{enumerate}
\end{lemma}

\vspace{-2mm} \subsection{The $\D$-chair} \vspace{-2mm}
The next definition classifies ten typed systems for the
$\D $-calculus: some of them already appeared (sometime with a
different notation) in the literature by the present authors.

\begin{definition}[$\D$-chair]\label{chair}\hfill\\
\begin{minipage}[b]{.65\textwidth}
 Ten typed systems $\DD{\TT}{\R}$ can be drawn pictorially in a
  $\D$-chair, where the arrows represent an inclusion relation. 
$\DD{CD}{\equiv}$ corresponds roughly to \cite{LR05,LR07} (in the expression
$M@\D$, $M$ is the essence of $\D$) and in its intersection
part to \cite{LFMTP17}; $\DD{CDS}{\equiv}$ corresponds roughly in its
intersection part to \cite{Dougherty-Liquori-10}, $\DD{BCD}{\equiv}$
corresponds in its intersection part to \cite{TTCS17},
$\DD{CD}{=_{\beta\eta}}$ corresponds in its intersection part to
\cite{APLAS16}. The other typed systems are basically new. 
The main properties of these systems are:
\begin{enumerate}
\item All the $\DD{\TT}{\equiv}$ systems enjoys the synchronous subject
  reduction property, the other systems also enjoy ordinary subject
  reduction (Th. \ref{SR1});
\item All the systems strongly normalize (Th. \ref{SN1});

\item All the systems correspond to the to original type assignment systems 
except $\DD{CD}{=_\beta}$,   $\DD{CDV}{=_\beta}$,  $\DD{CDV}{=_{\beta\eta}}$ and $\DD{BCD}{=_{\beta\eta}}$ (Th.   \ref{ISO1});
\item Type checking and type reconstruction are decidable for \mbox{all the
  systems, except $\DD{CDS}{=_\beta}$, $\DD{BCD}{=_\beta}$, and
  $\DD{BCD}{=_{\beta\eta}}$ (Th. \ref{TCTR1})}.
\end{enumerate}

%
\end{minipage}
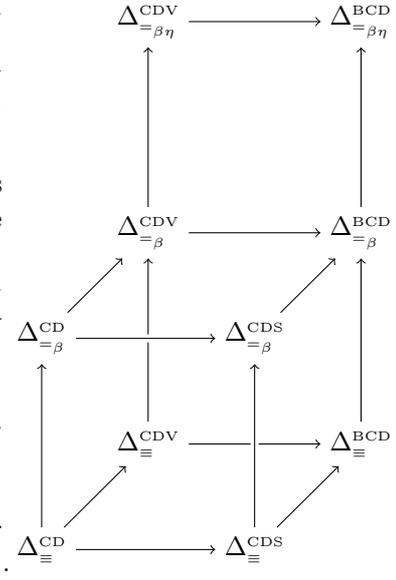
\begin{minipage}[b]{.35\textwidth}
  \begin{center}
    \begin{tikzpicture}[->,scale=1.4]
      \foreach \id/\name\x/\y in
      {a/\DD{CD}{\equiv}/0/0,b/\DD{CD}{=_\beta}
        /0/2,c/\DD{CDV}{\equiv}/1/1,d/\DD{CDV}{=_{\beta}}/1/3,
        e/\DD{CDS}{\equiv}/2/0,f/\DD{CDS}{=_\beta}/2/2,
        g/\DD{BCD}{\equiv}/3/1,h/\DD{BCD}{=_{\beta}}/3/3,
        i/\DD{CDV}{=_{\beta\eta}}/1/5,j/\DD{BCD}{=_{\beta\eta}}/3/5}
      \node (G-\id) at (\x,\y) {$\name$}; \foreach \from/\to in
      {c/g,e/f,e/g,f/h,g/h,a/b,a/c,b/d,c/d,a/e,b/f,d/h,d/i,h/j,i/j} {
        \draw[white, line width=3pt] (G-\from) -- (G-\to); \draw
        (G-\from) -- (G-\to); }
    \end{tikzpicture}
  \end{center}
  \end{minipage}

\end{definition}



\vspace{-3mm} \section{Examples}\label{exa} \vspace{-3mm}
This section shows  examples of typed derivations
$\DD{\TT}{\R}$ and highlights the corresponding type assignment
judgment in $\LTT$ they correspond to, in the sense that we have a
derivation $B\vdash^\TT_\R \D : \s$ and another derivation
$B\vdash^\TT_\cap \essence \D : \s$. The correspondence between
intersection typed systems $\DD{\TT}{\R}$ and intersection type
assignment $\LTT$ will be defined in Subsection
\ref{currychurch}.

\begin{example}[Polymorphic identity] 
 In all of the intersection type assignment systems $\LTT$ we can
  derive
  $\vdash^\TT_\cap \lambda x.x : (\s\to\s) \cap (\t\to\t)$
  A corresponding $\D$-term is:
  $\spair{\lambda x\of\s.x}{\lambda x\of\t.x}$
  that can be typed in all of the typed systems of the $\D$-chair
  as follows
  \begin{displaymath}
    \infer{\vdash^\TT_\R \spair{\lambda x\of\s.x}{\lambda x\of\t.x} :
      (\s\to\s) \cap (\t\to\t)}{
      \infer{\vdash^\TT_\R \lambda x\of\s.x : \s\to\s}
      {x\of\s\vdash^\TT_\R x:\s} &
      \infer{\vdash^\TT_\R \lambda x\of\t.x : \t\to\t}
      {x\of\t\vdash^\TT_\R x:\t} &
      \lambda x.x \mathrel{\R} \lambda x.x
    }
  \end{displaymath}
\end{example}

\begin{example}[Auto application] 
  In all of the intersection type assignment systems we can derive
  $\vdash^\TT_\cap \lambda x.x \at x :
    ((\s \to \t) \cap \s) \to \t 
   $
  A corresponding $\D$-term is:
  $\lambda x \of (\s \to \t) \cap \s. (\prl x) (\prr x)$
  that can be typed in all of the typed systems of the $\D$-chair
  as follows
  \begin{displaymath}
    \infer{\vdash^\TT_\R \lambda x \of (\s \to \t) \cap \s.
      (\prl x) (\prr x) : (\s \to \t) \cap \s \to \t}{
      \infer{x \of (\s \to \t) \cap \s \vdash^\TT_\R
        (\prl x) (\prr x) : \t}{
        \infer{x \of (\s \to \t) \cap \s \vdash^\TT_\R
          \prl x : \s \to \t}{
          x \of (\s \to \t) \cap \s \vdash^\TT_\R
          x : (\s \to \t) \cap \s
        } &
        \infer{x \of (\s \to \t) \cap \s \vdash^\TT_\R
          \prr x : \s}{
          x \of (\s \to \t) \cap \s \vdash^\TT_\R
          x : (\s \to \t) \cap \s
        }
      }
    }
  \end{displaymath}
\end{example}

\begin{example}[Some examples in  $\DD{CDS}{\R}$] 
  In $\LCDS$ we can derive
  $\vdash^\CDS_\cap (\lambda x.\lambda y. x) : \s\to\om\to\s$,
  and using this type assignment, we can derive
  $z\of\s\vdash^\CDS_\cap (\lambda x.\lambda y. x) \at z \at z : \s$.
  A corresponding $\D$-term is:
  $(\lambda x\of\s.\lambda y\of\om.x)\at z \at z^\om$
  that can be typed in $\DD{CDS}{\R}$ as follows
  \begin{displaymath}
    \infer{z\of\s\vdash^\CDS_\R
      (\lambda x\of\s.\lambda y\of\om.x)\at z \at z^\om : \s}{
      \infer{z\of\s\vdash^\CDS_\R
        (\lambda x\of\s.\lambda y\of\om.x)\at z : \om\to\s}{
        \infer{z\of\s\vdash^\CDS_\R
          \lambda x\of\s.\lambda y\of\om.x : \s\to\om\to\s}{
          \infer{z\of\s, x\of\s\vdash^\CDS_\R
            \lambda y\of\om.x : \om\to\s}{
            z\of\s, x\of\s,y\of\om \vdash^\CDS_\R x : \s
          }
        }
        &
        z\of\s\vdash^\CDS_\R z : \s
      }
      &
      \infer{z\of\s\vdash^\CDS_\R z^\om : \om}{
        z\of\s\vdash^\CDS_\R z : \s
        & \s\leq_\CDS\om
      }
    }
  \end{displaymath}
  As another example, we can also derive
  $\vdash^\CDS_\cap \lambda x.x : \s \to \s \cap \om$.
  A corresponding $\D$-term is:
  $ \lambda x\of\s. \spair{x}{x^\om} $
  that can be typed in $\DD{CDS}{\R}$ as follows
  \begin{displaymath}
    \infer{\vdash^\CDS_\R
      \lambda x\of\s. \spair{x}{x^\om} : \s\to\s\cap\om}{
      \infer{x\of\s\vdash^\CDS_\R
        \spair{x}{x^\om} : \s\cap\om}{
        x\of\s\vdash^\CDS_\R x : \s
        &
        \infer{x\of\s\vdash^\CDS_\R x^\om : \om}{
          x\of\s\vdash^\CDS_\R x : \s
          & \s\leq_\CDS\om
        }
        & x \mathrel{\R} x
      }
    }
  \end{displaymath}
\end{example}

\begin{example}[An example in $\DD{CDV}{\R}$]
  In $\LCDV$ we can prove the commutativity of intersection, \ie
  $ \vdash^\CDV_\cap \lambda x.x : \s \cap \t \to \t \cap \s $
  A corresponding $\D$-term is:
  $ \spair{\lambda x\of\s\cap\t.\prr x}
    {\lambda x\of\s\cap\t.\prl x}^{(\s\cap\t)\to(\t\cap\s)} $
  that can be typed in $\DD{CDV}{\R}$ as follows
  \begin{displaymath}
    \infer{\vdash^\CDS_\R\spair{\lambda x\of\s\cap\t.\prr x}
      {\lambda x\of\s\cap\t.\prl x}^{(\s\cap\t)\to(\t\cap\s)} :
      (\s\cap\t)\to(\t\cap\s)}{
      \infer{\vdash^\CDS_\R\spair{\lambda x\of\s\cap\t.\prr x}
        {\lambda x\of\s\cap\t.\prl x} :
        ((\s\cap\t)\to\t)\cap((\s\cap\t)\to\s)}{
        \infer{\vdash^\CDS_\R
          \lambda x\of\s\cap\t.\prr x : (\s\cap\t)\to\t}{
          \infer{x\of\s\cap\t\vdash^\CDS_\R \prr x : \t}{
            x\of\s\cap\t\vdash^\CDS_\R x : \s\cap\t
          }
        }
        &
        \infer{\vdash^\CDS_\R
          \lambda x\of\s\cap\t.\prl x : (\s\cap\t)\to\s}{
          \infer{x\of\s\cap\t\vdash^\CDS_\R \prl x : \s}{
            x\of\s\cap\t\vdash^\CDS_\R x : \s\cap\t
          }
        }
        &
        \lambda x.x \mathrel{\R} \lambda x.x
      }
      & \qquad \ast
    }
  \end{displaymath}
  where $\ast$ is
  $((\s\cap\t)\to\t)\cap((\s\cap\t)\to\s) \leq_\CDV
  (\s\cap\t)\to(\t\cap\s)$.
\end{example}

\begin{example}[Another polymorphic identity in $\DD{\TT}{=_\beta}$]
  In all the $\DD{\TT}{=_\beta}$ you can type this $\D$-term:
  $\spair{\lambda x\of\s.x}{(\lambda x\of\t{\to}\t.x)\at
      (\lambda x\of\t.x)} $
  The typing derivation is thus
  \begin{displaymath}
    \infer{\vdash^\TT_{=_\beta} \spair{\lambda x\of\s.x}
      {(\lambda x\of\t{\to}\t.x)\at(\lambda x\of\t.x)} :
      (\s\to\s) \cap (\t\to\t)}{
      \infer{\vdash^\TT_{=_\beta} \lambda x\of\s.x : \s\to\s}
      {x\of\s\vdash^\TT_{=_\beta} x:\s}
      \!\!\!&
      \infer{\vdash^\TT_{=_\beta}
        (\lambda x\of\t{\to}\t.x)\at(\lambda x\of\t.x) : \t\to\t}{
        \infer{\vdash^\TT_{=_\beta}
          \lambda x\of\t{\to}\t.x : (\t\to\t)\to(\t\to\t)}
        {x\of\t\to\t\vdash^\TT_{=_\beta} x:\t\to\t}
        \!\!\!&
        \infer{\vdash^\TT_{=_\beta} \lambda x\of\t.x : \t\to\t}
        {x\of\t\vdash^\TT_{=_\beta} x:\t}
      }
      \!\!\!&
      \hspace{-1.5cm}\lambda x.x =_\beta (\lambda x.x)\at(\lambda x.x)
    }
  \end{displaymath}
\end{example}

\begin{example}[Two examples in $\DD{BCD}{\equiv}$ and $\DD{BCD}{=_{\beta\eta}}$] 
  In $\LBCD$ we can can type any term, including the non-terminating
  term
  $\Omega \eqdef (\lambda x.x \at x) \at (\lambda x.x \at x) $
  More precisely, we have:
  $\vdash^\BCD_\cap \Omega : \om$
  A corresponding $\D$-term whose essence is $\Omega$ is:
  $ (\lambda x\of\om. x^{\om\to\om}\at x)\at
    (\lambda x\of\om. x^{\om\to\om}\at x)^\om $
  that can be typed in $\DD{BCD}{\R}$ as follows
  \begin{displaymath}
    \infer{\vdash^{\BCD}_{\R} (\lambda x\of\om. x^{\om\to\om}\at x)\at
      (\lambda x\of\om. x^{\om\to\om}\at x)^\om : \om}{
      \infer{\vdash^\BCD_\R
        \lambda x\of\om. x^{\om\to\om}\at x : \om\to\om}{\ast}
      &
      \infer{\vdash^\BCD_\R
        (\lambda x\of\om. x^{\om\to\om}\at x)^\om : \om}{
        \infer{\vdash^\BCD_\R
          \lambda x\of\om. x^{\om\to\om}\at x : \om\to\om}{\ast}
        & \om\to\om\leq_\BCD\om
      }
    }
  \end{displaymath}
  where $\ast$ is
  \begin{displaymath}
    \infer{x\of\om\vdash^\BCD_\R x^{\om\to\om}\at x : \om}{
      \infer{x\of\om\vdash^\BCD_\R x^{\om\to\om} : \om\to\om}{
        x\of\om\vdash^\BCD_\R x : \om
        & \om \leq_\BCD \om\to\om}
      & x\of\om\vdash^\BCD_\R x : \om
    }
  \end{displaymath}

  In $\LBCD$ we can type
  $ x \of \om\to\om \vdash^{\BCD}_\cap x : (\om\to\om) \cap
    (\s\to\om) $
  A corresponding $\D$-term whose essence is $x$ is:
  $\spair{x}{\lambda y\of \s. x \at y^\om} $
  that can be typed in $\DD{BCD}{=_{\beta\eta}}$ as follows
  \begin{displaymath}
    \infer{x \of \om\to\om \vdash^\BCD_{=_{\beta\eta}}
      \spair{x}{\lambda y\of \s. x \at y^\om} :
      (\om\to\om) \cap (\s\to\om)}{
      x \of \om\to\om \vdash^\BCD_{=_{\beta\eta}} x : \om\to\om
      & \mkern-100mu
      \infer{x \of \om\to\om \vdash^\BCD_{=_{\beta\eta}}
        \lambda y\of \s. x \at y^\om : \s\to\om}{
        \infer{x\of\om\to\om,y\of\s \vdash^\BCD_{=_{\beta\eta}}
          x\at y^\om : \om}{
          x\of\om\to\om,y\of\s \vdash^\BCD_{=_{\beta\eta}}
          x : \om\to\om &
          \infer{x\of\om\to\om,y\of\s \vdash^\BCD_{=_{\beta\eta}}
            y^\om : \om}{
            x\of\om\to\om,y\of\s \vdash^\BCD_{=_{\beta\eta}} y : \s
            & \s \leq \om
          }
        }
      }
      & \mkern-30mu
      x =_{\beta\eta} \lambda y. x \at y
    }
  \end{displaymath}
  Note that the $=_{\beta\eta}$ condition has an interesting loophole, as it is
  well-known that $\LBCD$ does not enjoy $=_\eta$-conversion property.
  Theorem \ref{SCI}(\ref{SOUND}) will show that we can construct a
  $\D$-term which does not correspond to any $\LBCD$ derivation.
\end{example}

\begin{example}[Pottinger] 
  The following examples can be typed in all the type theories of the
  $\D$-chair (we also display in square brackets the corresponding
  pure $\lambda$-terms typable in $\LTT$). These are encodings from
  the examples \ala\ Curry given by Pottinger in \cite{Pottinger-80}.
  \begin{displaymath}
    \begin{array}{ll}
      [\lambda x.\lambda y.x\at y] \quad 
      \vdash^\TT_\R \lambda x\of(\s \to \t)\cap (\s \to \r).
      \lambda y\of \s. \spair{(\prl x)\at y)}{(\prr x)\at y} :
      (\s \to \t)\cap (\s \to \r) \to \s \to \t\cap \r
      \\[2mm]
      [\lambda x.\lambda y.x\at y] \quad 
      \vdash^\TT_\R \lambda x\of \s \to \t \cap \r.
      \spair{\lambda y\of \s. \prl (x\at y)}
      {\lambda y\of \s .\prr (x\at y)} :
      (\s \to \t\cap \r) \to (\s \to \t)\cap (\s \to \r)
      \\[2mm]
      [\lambda x.\lambda y.x\at y] \quad 
      \vdash^\TT_\R \lambda x\of \s \to \r.\lambda y\of \s \cap \t.
      x\at(\prl y) : (\s \to \r) \to \s \cap \t\to \r
      \\[2mm]
      [\lambda x.\lambda y.x] \quad 
      \vdash^\TT_\R \lambda x\of \s \cap \t.\lambda y\of \s.
      \prr x : \s \cap \t \to \s \to \t
      \\[2mm]
      [\lambda x.\lambda y.x\at y \at y] \quad 
      \vdash^\TT_\R \lambda x\of\s\to\t\to\r.\lambda y\of\s\cap\t.
      x\at(\prl y)\at(\prr y) :
      (\s \to \t \to \r) \to \s \cap \t \to \r
      \\[2mm]
      [\lambda x.x] \quad 
      \vdash^\TT_\R \lambda x\of \s \cap \t.\prl x : \s \cap \t \to \s
      \\[2mm]
      [\lambda x.x] \quad 
      \vdash^\TT_\R \lambda x\of \s.\spair{x}{x} : \s \to \s \cap \s
      \\[2mm]
      [\lambda x.x] \quad 
      \vdash^\TT_\R \lambda x\of \s \cap (\t\cap \r).
      \spair{\spair{\prl x }{\prl \prr x}}{\prr \prr x} :
      \s \cap (\t\cap \r)\to (\s \cap \t)\cap \r
    \end{array}
  \end{displaymath}

  In the same paper, Pottinger lists some types that cannot be
  inhabited by any intersection type assignment ($\not\vdash^\TT_\cap$) in an empty context,
  namely: 
  $\s \to (\s \cap \t) \mbox{ and }   (\s \to \t) \to (\s \to \r) \to
     \s \to \t \cap \r \mbox{ and }  ((\s \cap \t) \to \r) \to \s \to \t \to \r$.
  %
  It is not difficult to verify that the above types cannot be
  inhabited by any of the type systems of the $\D$-chair because of
  the failure of the essence condition in the strong pair type rule.
\end{example}

\begin{example}[Intersection is not the conjunction operator] 
  This counter-example is from the corresponding counter-example \ala\
  Curry given by Hindley \cite{hindleylogic} and Ben-Yelles
  \cite{Ben-Yelles}.  The intersection type
   $(\s \to \s) \cap ((\s \to \t \to \r) \to (\s \to \t) \to
    \s \to \r)$
  where the left part of the intersection corresponds to the type for
  the combinator {\sf I} and the right part for the combinator {\sf S}
  cannot be assigned to a pure $\lambda$-term. Analogously, the same
  intersection type cannot be assigned to any $\D$-term.
\end{example}

\vspace{-4mm} \subsection{On synchronization and subject reduction} \vspace{-2mm}
For the typed systems $\DD{\TT}{\equiv}$, strong pairs have an
intrinsic notion of synchronization: some redexes need to be reduced
in a synchronous fashion unless we want to create meaningless
$\D$-terms that cannot be typed.  Consider the $\D$-term
$\spair{(\lambda x\of\s.x)\at y}{(\lambda x\of\s.x)\at y}$.  If we use
the $\red$ reduction relation, then the following reduction
paths are legal\\
$
  \spair{(\lambda x\of\s.x)\at y}{(\lambda x\of\s.x)\at y}
  \begin{array}{l}
    \nnearrow^{\beta} \spair{(\lambda x\of\s.x)\at y}{y}
    \ssearrow_{\beta}
    \\[2mm]
    \ssearrow_{\beta} \spair{y}{(\lambda x\of\s.x)\at y}
    \nnearrow^{\beta}
  \end{array}
  \spair{y}{y}.
$
More precisely, the first and second redexes are rewritten
asynchronously, thus they cannot be typed in any typed system
$\DD{\TT}{\equiv}$, because we fail to check the left and the right
part of the strong pair to be the same: the
$\red^\parallel$ reduction relation prevents this loophole
and allows to type all redexes. In summary,
$\red^\parallel$ can be thought of as the natural
reduction relation for the typed systems $\DD{\TT}{\equiv}$.



\vspace{-3mm} \section{Metatheory of $\DD{\TT}{\R}$}\label{meta} \vspace{-3mm}

\subsection{General properties} \vspace{-2mm}
Unless specified, all properties applies to the intersection typed
systems $\DD{\TT}{\R}$. For lack of space all  proofs are omitted: the interested 
reader can found more technical details in the Appendix.
The Church-Rosser property is proved using the technique of
Takahashi \cite{takahashi1995parallel}. The parallel reduction
semantics extends Definition \ref{red} and it is inductively defined
as follows.
\begin{definition}[Parallel reduction semantics] \label{taka}
  \begin{displaymath}
    \begin{array}[c]{r@{\quad}c@{\quad}lll}
      x & \taka
      & x &
      \multicolumn{2}{l}
      {and  \qquad \qquad u_\D   \taka   u_\D}
      \\[2mm]
      \D^\s
        & \taka & (\D')^\s & \IF \D\taka\D'
      \\[2mm]
      \D_1\at\D_2
        & \taka & \D'_1\at\D'_2
          & \IF \D_1 \taka \D'_1 & \AND \D_2 \taka \D'_2
      \\[2mm]
      \lambda x\of\s.\D
        & \taka & \lambda x\of\s.\D' & \IF \D\taka\D'
      \\[2mm]
      (\lambda x\of\s.\D_1)\at\D_2
        & \taka & \subst{\D'_1}{x}{\D'_2}
          & \IF \D_1 \taka \D'_1 & \AND \D_2 \taka \D'_2
      \\[2mm]
      \spair{\D_1}{\D_2}
        & \taka & \spair{\D'_1}{\D'_2}
          & \IF \D_1 \taka \D'_1 & \AND \D_2 \taka \D'_2
      \\[2mm]
      \pri\D
        & \taka & \pri\D'
          & \IF \D \taka \D' & \AND i \in\{1,2\}
      \\[2mm]
      \pri\spair{\D_1}{\D_2}
        & \taka & \D'_i
          & \IF \D_i \taka \D'_i & \AND i \in\{1,2\}
      \\[2mm]
    \end{array}
  \end{displaymath}
\end{definition}
Intuitively, $\D \taka \D'$ means that $\D'$ is obtained from $\D$ by
simultaneous contraction of some $\beta\pri$-redexes possibly
overlapping each other.  Church-Rosser can be achieved by proving a
stronger statement, namely
  $\D \taka \D' \quad  \IMPLIES \quad \D' \taka \D^\ast \label{ss}$
where $\D^\ast$ is a $\D$-term determined by $\D$ and independent from
$\D'$. The statement (\ref{ss}) is satisfied by the term $\D^\ast$
which is obtained from $\D$ by contracting all the redexes existing in
$\D$ simultaneously.

\begin{definition}[The map $\_^\ast$]\label{ast} \hfill
  \begin{displaymath}
    \begin{array}{rcl@{\qquad \qquad \qquad}rcl}
      x^\ast & \eqdef & x
     &
      u_\D^\ast & \eqdef & u_\D
      \\[2mm]
      (\D^\s)^\ast& \eqdef &(\D^\ast)^\s
      &
      \spair{\D_1}{\D_2}^\ast & \eqdef & \spair{\D^\ast_1}{\D^\ast_2}
      \\[2mm]
      (\lambda x\of\s.\D)^\ast & \eqdef & \lambda x\of\s. \D^\ast
      &
      ((\lambda x\of\s.\D_1)\at\D_2)^\ast & \eqdef & \subst{\D^\ast_1}{x}{\D^\ast_2}
      \\[2mm]
       (\D_1\at\D_2)^\ast  & \eqdef &   \D^\ast_1\at\D^\ast_2 &   \multicolumn{3}{l}{\mbox{if $\D_1\at\D_2$ is not a $\beta$-redex}}
      \\[2mm]
      (\lambda x\of\s.\D\at x)^\ast  & \eqdef &  \D^\ast & \multicolumn{3}{l}{\mbox{if $x \not\in {\rm fv}(\D)$}}
       \\[2mm]
      (\pri\D)^\ast  & \eqdef &   \pri\D^\ast & \multicolumn{3}{l}{ \mbox{if $\D$ is not a strong pair}}
      \\[2mm]
    \end{array}
  \end{displaymath}
\end{definition}
%
%

Now we have to prove the Church-Rosser property for the parallel
reduction.
\begin{lemma}[Confluence property for $\taka$]\label{lem:takaconfl} 
  If $\D \taka \D'$, then $\D' \taka \D^\ast$.
\end{lemma}

The Church-Rosser property follows.
\begin{theorem}[Confluence]
  If $\D_1 \multired \D_2$ and
  $\D_1 \multired \D_3$, then there exists $\D_4$ such
  that $\D_2 \multired \D_4$ and
  $\D_3 \multired \D_4$.
\end{theorem}

The next lemma says that all type derivations for $\D$ have an unique
type.
\begin{lemma}[Unicity of typing]\label{UNIQUE}
  If $B \vdash^\TT_\R \D : \s$, then $\s$ is unique.
\end{lemma}
The next theorem states that all the $\DD{\TT}{\equiv}$ typed systems
preserve synchronous $\beta\pri$-reduction, and all the
$\DD{\TT}{=_{\beta}}$ and $\DD{\TT}{=_{\beta\eta}}$ typed systems
preserve $\beta\pri$-reduction.
\begin{theorem}[Subject reduction for $\beta\pri$]\hfill\label{SR1}
  \begin{enumerate}
  \item If $B \vdash^\TT_\equiv \D_1 : \s$ and
    $\D_1 \red^\parallel \D_2$, then
    $B \vdash^\TT_\equiv \D_2 : \s$;
  \item for $\R \in \{=_\beta,=_{\beta\eta}\}$, if
    $B \vdash^\TT_\R \D_1 : \s$ and $\D_1 \red \D_2$, then
    $B \vdash^\TT_\R \D_2 : \s$.
  \end{enumerate}
\end{theorem}

The next theorem states that some of the typed systems on the back of
the $\D$-chair preserve $\eta$-reduction.
\begin{theorem}[Subject reduction for $\eta$ for
  $\CDV,\BCD$]\label{SRETA}
  Let $\TT \in \{\CDS,\BCD\}$. If
  $B \vdash^\TT_{=_{\beta\eta}} \D_1 : \s$ and $\D_1 \red_\eta \D_2$,
  then $B \vdash^\TT_{=_{\beta\eta}} \D_2 :\s$.
\end{theorem}

\vspace{-6mm} \subsection{Strong normalization} \vspace{-2mm}
The idea of the strong normalization proof is to embed typable terms
of the $\Delta$-calculus into Church-style terms of a target system,
which is the simply-typed $\lambda$-calculus with pairs, in a
structure-preserving way (and forgetting all the essence
side-conditions). The translation is sufficiently faithful so as to
preserve the number of reductions, and so strong normalization for the
$\D$-calculus follows from strong normalization for simply-typed
$\lambda$-calculus with pairs. A similar technique has been used in \cite{LF} 
to prove the strong normalization property of LF 
and in \cite{bucciarelli} to prove the strong normalization property of a subset of $\LCD$.

The target system has one atomic type called $\circ$, a special
constant term $u_\circ$ of type $\circ$ and an infinite number of
constants $c_{\s}$ of type $\s$ for any type of the target
system. We denote by $B \vdash_\times M : \s$ a typing judgment in
the target system.

\begin{definition}[Forgetful mapping] 
  \begin{itemize}
  \itemsep = -1mm
  \item On intersection types.
    \[
        \map{{\tt a}_i} ~ \eqdef ~ \circ \quad
                                   \forall {\tt a}_i \in\Adef
                                   \qquad \mbox{and} \qquad 
        \map{\s \cap \t} ~\eqdef ~ \map{\s} \times \map{\t}
       \qquad \mbox{and} \qquad 
        \map{\s \to \t} ~ \eqdef ~ \map{\s} \to \map{\t}
    \]
  \item On $\D$-terms.
    \begin{displaymath}
      \begin{array}{rcl@{\qquad}rcl}
        \map{x}_B & \eqdef & x 
        &
        \map{u_\D}_B & \eqdef & u_\circ 
        \\[2mm]
        \map{\lambda x\of\s.\D}_B & \eqdef
                           &
                             \lambda x. \map{\D}_{B,x\of\s}  
       &                       
        \map{\D_1\at\D_2}_B & \eqdef
                           & \map{\D_1}_B \at \map{\D_2}_B 
       \\[2mm]
        \map{\spair{\D_1}{\D_2}}_B & \eqdef
                           & \pair{\map{\D_1}_B}{\map{\D_2}_B} 
       &                    
        \map{\pri \D}_B & \eqdef & \pri \map{\D}_B 
        \\[2mm]
        \map{\D^\t}_B & \eqdef & c_{\map{\s}\to\map{\t}} \at \map{\D}_B
                                 \quad\mbox{if $B\vdash^\TT_\R \D:\s$}
      \end{array}
    \end{displaymath}
    \item     The map can be easily extended to basis $B$.

  \end{itemize}
\end{definition}
%

\begin{theorem}[Strong normalization]\label{SN1}
  If $B \vdash^\TT_\R \D : \s$, then $\D$ is strongly normalizing.
\end{theorem}


\section{Typed systems \ala\ Church \vs\ type assignment systems \ala\
  Curry}\label{meta-2} \vspace{-3mm}
\subsection{\mbox{Relation between type assignment systems $\LTT$ and typed
  systems $\DD{\TT}{\R}$}}\label{currychurch} \vspace{-2mm}
It is interesting to state some relations between type assignment
systems \ala\ Church and typed systems \ala\ Curry. An interesting
property is the one of isomorphism, namely the fact that whenever we
assign a type $\s$ to a pure $\lambda$-term $M$, the same type can be
assigned to a $\D$-term such that the essence of $\D$ is
$M$. Conversely, for every assignment of $\s$ to a $\D$-term, a valid
type assignment judgment of the same type for the essence of $\D$ can
be derived.
Soundness, completeness and isomorphism between intersection typed
systems for the $\D$-calculus and the corresponding intersection type
assignment systems for the $\lambda$-calculus are defined as follows.
\begin{definition}[Soundness, completeness and
  isomorphism]\label{SCI}
  Let $\DD{\TT}{\R}$ and $\LTT$.
  \begin{enumerate}
  \item \label{SOUND}(Soundness, $\DD{\TT}{\R} \triangleleft
    \LTT$). $B \vdash^\TT_\R \D : \s$ implies
    $B \vdash^\TT_\cap \essence{\D} : \s$;
  \item (Completeness, $\DD{\TT}{\R} \triangleright \LTT$).
    $B \vdash^\TT_\cap M :\s $ implies there exists $\D$ such that
    $M\equiv\essence{\D}$ and $B \vdash^\TT_\R \D : \s$;
  \item (Isomorphism, $\DD{\TT}{\R} \sim \LTT$).
    $\DD{\TT}{\R} \triangleright \LTT$ and
    $\DD{\TT}{\R} \triangleleft \LTT$.
  \end{enumerate}
\end{definition}

\mbox{The following properties and relations between typed and type
assignment systems can be verified.}

\begin{figure}[t]
 \[
    \begin{array}{|l|c|c|}
      \hline\\[-4.5mm]
      \DD{\TT}{\R} & \DD{\TT}{\R} \triangleleft \LTT
      & \DD{\TT}{\R} \triangleright \LTT
      \\[.5mm]\hline
      \DD{CD}{\equiv} & \surd & \surd
      \\[.5mm]
      \DD{CDV}{\equiv} & \surd & \surd
      \\[.5mm]
      \DD{CDS}{\equiv} & \surd & \surd
      \\[.5mm]
      \DD{BCD}{\equiv} & \surd & \surd
      \\[.5mm]\hline
      \DD{CD}{=_\beta} & \times & \surd
      \\[.5mm]
      \DD{CDV}{=_\beta} & \times & \surd
      \\[.5mm]
      \DD{CDS}{=_\beta} & \surd & \surd
      \\[.5mm]
      \DD{BCD}{=_\beta} & \surd & \surd
      \\[.5mm]\hline
      \DD{CDV}{=_{\beta\eta}} & \times & \surd
      \\[.5mm]
      \DD{BCD}{=_{\beta\eta}} & \times & \surd
      \\[.5mm]\hline
    \end{array}
    \qquad     \qquad
    \begin{array}{|l|c|}
      \hline\\[-4.5mm]
      \DD{\TT}{\R} & {\rm TC/TR}
      \\[.5mm]\hline
      \DD{CD}{\equiv}
                   & \surd
      \\[.5mm]
      \DD{CDV}{\equiv}
                   & \surd
      \\[.5mm]
      \DD{CDS}{\equiv}
                   & \surd
      \\[.5mm]
      \DD{BCD}{\equiv}
                   & \surd
      \\[.5mm]\hline
      \DD{CD}{=_\beta}
                   & \surd
      \\[.5mm]
      \DD{CDV}{=_\beta}
                   & \surd
      \\[.5mm]
      \DD{CDS}{=_\beta}
                   & \times
      \\[.5mm]
      \DD{BCD}{=_\beta}
                   & \times
      \\[.5mm]\hline
      \DD{CDV}{=_{\beta\eta}}
                   & \surd
      \\[.5mm]
      \DD{BCD}{=_{\beta\eta}}
                   & \times
      \\[.5mm]\hline
    \end{array}
 \qquad     \qquad
  \begin{array}{|l|l|l|}
      \hline\\[-4.5mm]
      \mbox{Source} & \mbox{Target}
      \\[.5mm]\hline
      \DD{CD}{\equiv}
                    &
                      \DD{CD}{=_\beta}
      \\[.5mm]
      \DD{CDV}{\equiv}
                    &
                      \DD{CDV}{=_{\beta\eta}}
      \\[.5mm]
      \DD{CDS}{\equiv}
                    &
                      \DD{CDS}{=_{\beta}}
      \\[.5mm]
      \DD{BCD}{\equiv}
                    &
                      \DD{BCD}{=_{\beta\eta}}
      \\[.5mm]\hline
      \DD{CD}{=_\beta}
                    &
                      \DD{CD}{=_\beta} 
      \\[.5mm]
      \DD{CDV}{=_\beta}
                    &
                      \DD{CDV}{=_{\beta\eta}}
      \\[.5mm]
      \DD{CDS}{=_\beta}
                    &
                      \DD{CDS}{=_{\beta}} 
      \\[.5mm]
      \DD{BCD}{=_\beta}
                    &
                      \DD{BCD}{=_{\beta\eta}}
      \\[.5mm]\hline
      \DD{CDV}{=_{\beta\eta}}
                    &
                      \DD{CDV}{=_{\beta\eta}} 
      \\[.5mm]
      \DD{BCD}{=_{\beta\eta}}
                    &
                      \DD{BCD}{=_{\beta\eta}} 
      \\[.5mm]\hline
    \end{array}
  \]
\caption{On the left: Soundness, completeness, isomorphism. On the center: type checking/reconstruction. On the right: source and target languages of the translation}\label{ALL}
\end{figure}

\begin{theorem}[Soundness, completeness and isomorphism]\label{ISO1}
  The following properties (left of Figure \ref{ALL}) between $\D$-calculi and type assignment
  systems $\LTT$ can be verified.
\end{theorem}

The last theorem characterizes the class of strongly normalizing
$\D$-terms.
\begin{theorem}[Characterization] 
  Every strongly normalizing $\lambda$-term can be type-annotated so
  as to be the essence of a typable $\D$-term.
\end{theorem}

We can finally state decidability of type checking (TC) and type
reconstruction (TR).
\begin{theorem}[Decidability of type checking and type reconstruction]\label{TCTR1}
  Figure \ref{ALL} (in the center) list decidability of type checking and type reconstruction.
\end{theorem}

\vspace{-4mm} \subsection{Subtyping and explicit coercions}\label{tannen} \vspace{-2mm}
The typing rule $(\leq_\TT)$ in the general typed system introduces
type coercions: once a type coercion is introduced, it cannot be
eliminated, so {\it de facto} freezing a $\D$-term inside an explicit
coercion.  Tannen \etal\ \cite{TannenCGS91} showed a translation of a
judgment derivation from a ``Source" system with subtyping (Cardelli's
Fun \cite{CardFun}) into an ``equivalent'' judgment derivation in a
``Target''  system without subtyping (Girard system F with records and
recursion).  In the same spirit, we present a translation that removes
all explicit coercions.  Intuitively, the translation proceeds as
follows: every derivation ending with rule $(\leq_\TT)$ is translated into 
the following (coercion-free) derivation, \ie\ \\
$
  \infer[({\to}E)] {B \vdash^\TT_{\R'} \trans{\s \leq_\TT \t} \at
    \trans{\D}_B : \t} {B \vdash^\TT_{\R'} \trans{\s \leq_\TT \t} : \s
    \to \t & B \vdash^\TT_{\R'} \trans{\D}_B : \s}
$
\\ where $\R'$ is a suitable relation such that $\R \sqsubseteq
\R'$. Note that changing of the type theory is necessary to guarantee
well-typedness in the translation of strong pairs. Summarizing, we
provide a type preserving translation of a $\D$-term into a
coercion-free $\D$-term such that
$\essence{\D} =_{\beta\eta} \essence{\D'}$.
%
%
The following example illustrates some trivial compilations of axioms and rule schemes of Figure \ref{subtype}.
\begin{example}[Translation of axioms and rule schemes of Figure \ref{subtype}]\hfill
  \begin{enumerate}
  \item[(refl)\!\!] the judgment $x \of \s \vdash^\TT_\R  \spair{x}{x^\s} : \s \cap \s$ is
    translated to a coercion-free judgment\\
 $    x \of \s \vdash^\TT_{=_\beta} \spair{x}{(\lambda y\of \s.y) \at x} : \s \cap \s $
  \item[(incl)\!\!] the judgment
    $x \of \s \cap \t \vdash^\TT_\R \spair{x}{x^\t} : (\s \cap \t) \cap \t$ is
    translated to a coercion-free judgment\\
    $x \of \s \cap \t \vdash^\TT_{=_\beta} \spair{x}{(\lambda y\of
        \s \cap \t. \prr y) \at x} : (\s \cap \t) \cap \t $
  \item[(glb)\!\!] the judgment
    $x \of \s \vdash^\TT_\R \spair{x}{x^{\s\cap\s}} : \s \cap (\s
    \cap\s)$ is translated to a coercion-free judgment\\
    $ x \of \s \vdash^\TT_{=_\beta} \spair{x}{(\lambda y\of
        \s. \spair{y}{y}) \at x} : \s \cap (\s \cap \s) $
  \item[$(\om_{top})$\!\!] the judgment
    $x\of\s \vdash^\TT_\R \spair{x}{x^\om} : \s \cap \om $ is translated to a
    coercion-free judgment\\
    $ x\of\s \vdash^\TT_{=_\beta} \spair{x}{(\lambda y\of\s.u_y)\at x} : \s \cap \om $
  \item[$(\om_{\to})$\!\!] the judgment
    $x\of\om \vdash^\TT_\R \spair{x}{x^{\s\to\om}} : \om \cap (\s\to\om)$ is translated to
    a coercion-free judgment\\
    $ x\of\om \vdash^\TT_{=_{\beta\eta}} \spair{x}{(\lambda f\of\om.\lambda
      y\of\s. u_{(f\at y)})\at x} : \om \cap (\s\to\om) $
  \item[$({\to}\cap)$\!\!] the judgment
    $x\of(\s\to\t)\cap(\s\to\r) \vdash^\TT_\R x^{\s\to\t\cap\r} :
    \s\to \t\cap\r$ is translated to a coercion-free judgment\\
    $ x\of(\s\to\t)\cap(\s\to\r) \vdash^\TT_{=_{\beta\eta}} (\lambda
      f\of(\s\to\t)\cap(\s\to\r).\lambda y\of\s.\spair{(\prl f)\at
        y}{(\prr f)\at y})\at x : \s\to \t\cap\r $
  \item[$(\to)$\!\!] the judgment
    $x\of\s \to \t\cap\r \vdash^\TT_\R \spair{x}{x^{\s\cap \r \to\t}}: (\s \to \t\cap\r) \cap (\s\cap \r
    \to\t)$ is translated to a coercion-free judgment\\
    $ x\of\s\to \t\cap\r \vdash^\TT_{=_{\beta\eta}} \spair{x}{(\lambda
      f\of\s\to \t\cap\r. \lambda y\of\s\cap\r.\prl (f\at (\prl
      y)))\at x} : (\s \to \t\cap\r) \cap (\s\cap \r  \to\t) $
  \item[(trans)\!\!\!] the judgment
    $x \of \s \vdash^\TT_\R \spair{x}{(x^\om)^{\s \to\om}} : \s \cap (\s \to \om)$ is
    translated to a coercion-free judgment\\
    $ x \of \s \vdash^\TT_{=_{\beta\eta}} \spair{x}{(\lambda f\of \om.\lambda
      y\of \s.u_{(f \at y)}) \at ((\lambda y\of\s.u_y) \at x)} : \s \cap (\s \to \om) $
  \end{enumerate}
\end{example}

The next definition introduces two maps translating subtype judgments
into explicit coercions functions and $\D$-terms into coercion-free
$\Delta$-terms.
\begin{definition}[Translations $\trans{-}$ and $\trans{-}_B$]\hfill
  \begin{enumerate}
  \item The minimal type theory $\leqmin$ and the extra axioms and
    schemes are translated as follows.
    \[
      \begin{array}[c]{lrcl} 
      {\rm (refl)}
        &
        \trans{\s \leq_\TT \s}
        & \eqdef
        & \vdash^\TT_{=_\beta}\lambda x\of\s.x : \s\to\s
        \\[2mm]
        {\rm (incl_1)}
        &
          \trans{\s \cap \t \leq_\TT \s}
        & \eqdef
        & \vdash^\TT_{=_\beta}\lambda x\of\s\cap\t.\prl x :
          \s \cap\t \to \s
        \\[2mm]
        {\rm (incl_2)}
        &
          \trans{\s \cap \t \leq_\TT \t}
        & \eqdef
        & \vdash^\TT_{=_\beta}\lambda x\of\s\cap\t.\prr x :
          \s \cap\t \to \t
        \\[2mm]
        {\rm (glb)}
        &
          \trans{\dfrac{\r \leq_\TT \s \quad \r \leq_\TT \t}
          {\r \leq_\TT \s \cap \t}}
        & \eqdef
        & {\vdash^\TT_{=_\beta}\lambda x\of\r.
          \spair{\trans{\r \leq_\TT \s} x}{\trans{\r \leq_\TT \t} x} :
          \r \to \s \cap \t}
        \\[4mm]
        {\rm (trans)}
        &
          \trans{\dfrac{\s \leq_\TT \t \quad \t \leq_\TT \r}{\s \leq_\TT \r}}
        & \eqdef
        & {\vdash^\TT_{=_\beta} \lambda x\of\s. \trans{\t \leq_\TT \r}
          \at (\trans{\s \leq_\TT \t} \at x) : \s \to \r}
        \\[4mm]
        (\om_{top})
        &
          \trans{\s \leq_\TT\om}
        &\eqdef
        & \vdash^\TT_{=_\beta}\lambda x\of\s.u_x : \s\to\om
        \\[2mm]
        (\om_\to)
        &
          \trans{\om \leq_\TT \s \to \om}
        & \eqdef
        & \vdash^\TT_{=_{\beta\eta}}\lambda f\of\om.\lambda x\of \s.
          u_{(f\at x)} : \om\to(\s\to\om)
        \\[2mm]
        \multicolumn{4}{l}{\rm Let~\xi_1 \eqdef (\s \to \t) \cap (\s \to \r)
        ~and~\xi_2 \eqdef \s \to \t \cap \r}
        \\[2mm]
        ({\to}\cap)
        &
          \trans{\xi_1 \leq_\TT \xi_2}
        &\eqdef
        & \vdash^\TT_{=_{\beta\eta}} \lambda f\of\xi_1.\lambda x\of\s.
          \spair{(\prl f)\at x}{(\prr f)\at x} : \xi_1 \to \xi_2
        \\[4mm]
        \multicolumn{4}{l}{\rm Let~\xi_1 \eqdef \s_1 \to \t_1
        ~and~\xi_2 \eqdef \s_2 \to \t_2}
        \\[2mm]
        (\to)
        &
          \trans{\dfrac{\s_2\leq_\TT \s_1 \quad \t_1 \leq_\TT\t_2}
          {\s_1 \to\t_1\leq_\TT\s_2 \to\t_2}}
        & \eqdef
        & \vdash^\TT_{=_{\beta\eta}} \lambda f\of\xi_1.\lambda x\of\s_2.
          \trans{\t_1\leq_\TT \t_2}
         \at  (f\at(\trans{\s_2\leq_\TT \s_1} x)) : \xi_1 \to \xi_2
        \\[4mm]
      \end{array}
    \]
  \item The translation $\trans{-}_B$ is defined on $\D$ as follows.
    $$\begin{array}{rcl@{\qquad}rcl}
      \trans{u_\D}_B
       &\eqdef &
       u_{\trans{\D}_B}
       &
      \trans{x}_B
       &\eqdef&  
      x
      \\[2mm]
      \trans{\lambda x\of\s.\D}_B
       &\eqdef&
       \lambda x\of\s.\trans{\D}_{B,x\of \s}
      &
      \trans{\D_1 \at \D_2}_B
      & \eqdef &
      \trans{\D_1}_B \at \trans{\D_2}_B
      \\[2mm]
      \trans{\spair{\D_1}{\D_2}}_B
      & \eqdef
      & \spair{\trans{\D_1}_B}{\trans{\D_2}_B}
      &
      \trans{\pri \D}_B
      & \eqdef &
      \pri \trans{\D}_B \qquad i \in \{1,2\}
      \\[2mm]
      \trans{\D^\t}_B
      & \eqdef
      & \trans{\s \leq_\TT \t} \at \trans{\D}_B 
      &\multicolumn{3}{l}{  \IF B\vdash^\TT_\R \D :\s.}
    \end{array}
    $$
  \end{enumerate}
\end{definition}

By looking at the above translation functions we can see that if
$B \vdash^\TT_\R \D : \s$, then $\trans{\D}_B$ is defined and it is
coercion-free.
The following lemma states that a coercion function is always typable
in $\DD{\TT}{=_{\beta\eta}}$, that it is essentially the identity and
that, without using the rule schemes $({\to}\cap)$, $(\om_{\to})$, and $(\to)$
the translation can even be derivable in $\DD{\TT}{=_\beta}$.

\begin{lemma}[Essence of a coercion is an identity]\label{coerce}
  \begin{enumerate}
  \item If $\s \leq_\TT \t$, then
    $\vdash^\TT_{=_{\beta\eta}} \trans{\s\leq_\TT \t} : \s \to \t$ and
    $\essence{\trans{\s\leq_\TT \t}} =_{\beta\eta} \lambda x.x$;
  \item If $\s \leq_\TT \t$ without using the rule schemes
    $({\to}\cap)$, $(\om_{\to})$, and $(\to)$, then
    $\vdash^\TT_{=_\beta} \trans{\s\leq_\TT \t} : \s \to \t$ and
    $\essence{\trans{\s\leq_\TT \t}} =_\beta \lambda x.x$.
  \end{enumerate}
\end{lemma}

We can now prove the coherence of the translation as follows.
\begin{theorem}[Coherence] 
  If $B \vdash^\TT_\R \D : \s$, then
  $B \vdash^\TT_{\R'} \trans{\D}_B : \s$ and
  $\essence{\trans{\D}_B} \mathrel{\R'} \essence{\D}$, where
  $\DD{\TT}{\R}$ and $\DD{\TT}{\R'}$ are respectively the source and
  target intersection typed systems given in Figure \ref{ALL} (right part).
\end{theorem}
%


\bibliography{inter_biblio}

\appendix

\section{Metatheory of $\DD{\TT}{\R}$}\label{meta-1}
\subsection{General properties}
Unless specified, all properties applies to the intersection typed
systems $\DD{\TT}{\R}$.

The Church-Rosser property is proved using the technique of
Takahashi \cite{takahashi1995parallel}. The parallel reduction
semantics extends Definition \ref{red} and it is inductively defined
as follows.
\begin{definition}[Parallel reduction semantics] \label{taka}
  \begin{displaymath}
    \begin{array}[c]{r@{\quad}c@{\quad}lll }
      x & \taka
      & x &
      \\[2mm]
      u_\D & \taka & u_\D &
      \\[2mm]
      \D^\s
        & \taka & (\D')^\s & \IF \D\taka\D'
      \\[2mm]
      \D_1\at\D_2
        & \taka & \D'_1\at\D'_2
          & \IF \D_1 \taka \D'_1 & \AND \D_2 \taka \D'_2
      \\[2mm]
      \lambda x\of\s.\D
        & \taka & \lambda x\of\s.\D' & \IF \D\taka\D'
      \\[2mm]
      (\lambda x\of\s.\D_1)\at\D_2
        & \taka & \subst{\D'_1}{x}{\D'_2}
          & \IF \D_1 \taka \D'_1 & \AND \D_2 \taka \D'_2
      \\[2mm]
      \spair{\D_1}{\D_2}
        & \taka & \spair{\D'_1}{\D'_2}
          & \IF \D_1 \taka \D'_1 & \AND \D_2 \taka \D'_2
      \\[2mm]
      \pri\D
        & \taka & \pri\D'
          & \IF \D \taka \D' & \AND i \in\{1,2\}
      \\[2mm]
      \pri\spair{\D_1}{\D_2}
        & \taka & \D'_i
          & \IF \D_i \taka \D'_i & \AND i \in\{1,2\}
      \\[2mm]
    \end{array}
  \end{displaymath}
\end{definition}
Intuitively, $\D \taka \D'$ means that $\D'$ is obtained from $\D$ by
simultaneous contraction of some $\beta\pri$-redexes possibly
overlapping each other.  Church-Rosser can be achieved by proving a
stronger statement, namely
\begin{eqnarray}
  \D \taka \D' & \IMPLIES & \D' \taka \D^\ast \label{ss}
\end{eqnarray}
where $\D^\ast$ is a $\D$-term determined by $\D$ and independent from
$\D'$. The statement (\ref{ss}) is satisfied by the term $\D^\ast$
which is obtained from $\D$ by contracting all the redexes existing in
$\D$ simultaneously.

\begin{definition}[The map $\_^\ast$]\label{ast} \hfill
  \begin{displaymath}
    \begin{array}{rcll}
      x^\ast & \eqdef & x
      \\[2mm]
      u_\D^\ast & \eqdef & u_\D
      \\[2mm]
      (\D^\s)^\ast& \eqdef &(\D^\ast)^\s
      \\[2mm]
      \spair{\D_1}{\D_2}^\ast & \eqdef & \spair{\D^\ast_1}{\D^\ast_2}
      \\[2mm]
      (\lambda x\of\s.\D)^\ast & \eqdef & \lambda x\of\s. \D^\ast
      \\[2mm]
      (\D_1\at\D_2)^\ast & \eqdef & \D^\ast_1\at\D^\ast_2 & \mbox{if $\D_1\at\D_2$ is not a $\beta$-redex}
      \\[2mm]
      ((\lambda x\of\s.\D_1)\at\D_2)^\ast & \eqdef & \subst{\D^\ast_1}{x}{\D^\ast_2}
      \\[2mm]
      (\pri\D)^\ast & \eqdef & \pri\D^\ast & \mbox{if $\D$ is not a strong pair}
      \\[2mm]
      (\pri\spair{\D_1}{\D_2})^\ast & \eqdef & \D^\ast_i & i \in\{1,2\}
    \end{array}
  \end{displaymath}
\end{definition}

The next technical lemma will be useful in showing that Church-Rosser
for $\multired$ can be inherited from Church-Rosser for
$\taka$.

\begin{lemma}\label{lem:takaprop}\hfill
  \begin{enumerate}
  \item If $\D_1 \red \D'_1$, then $\D_1 \taka \D'_1$;
  \item if $\D_1 \taka \D'_1$, then
    $\D_1 \multired \D'_1$;
  \item if $\D_1 \taka \D'_1$ and $\D_2 \taka \D'_2$, then
    $\subst{\D_1}{x}{\D_2} \taka \subst{\D'_1}{x}{\D'_2}$;
  \item $\D_1 \taka \D^\ast_1$.
  \end{enumerate}
  \begin{proof}
    {\it ($\it 1$)} can be proved by induction on the context of the
    redexes, while {\it ($\it 2$)}, {\it ($\it 3$)}, and {\it
      ($\it 4$)} can be proved by induction on the structure of
    $\D_1$.
  \end{proof}
\end{lemma}

Now we have to prove the Church-Rosser property for the parallel
reduction.
\begin{lemma}[Confluence property for $\taka$]\label{lem:takaconfl}\hfill\\
  If $\D \taka \D'$, then $\D' \taka \D^\ast$.
  \begin{proof}
    By induction on the shape of $\D$.
    \begin{itemize}
    \item if $\D \equiv x$, then
      $\D' \equiv x \taka x \equiv \D^\ast$;
    \item if $\D \equiv u_\D$, then
      $\D' \equiv u_\D \taka u_\D \equiv \D^\ast$;
    \item if $\D \equiv \D_1^\s$, then, for some $\D'_1$, we have that
      $\D_1 \taka \D'_1$ and $\D' \equiv (\D'_1)^\s$, therefore, by
      induction hypothesis, $\D' \taka (\D^\ast_1)^\s \equiv \D^\ast$;
    \item if $\D \equiv \spair{\D_1}{\D_2}$, then, for some $\D'_1$
      and $\D'_2$, we have that $\D_1 \taka \D'_1$, $\D_2 \taka \D'_2$
      and $\D' \equiv \spair{\D'_1}{\D'_2}$. By induction hypothesis,
      $\D' \taka \spair{\D^\ast_1}{\D^\ast_2} \equiv \D^\ast$;
    \item if $\D \equiv \lambda x\of\s. \D_1$, then, for some $\D'_1$,
      we have that $\D_1 \taka \D'_1$ and
      $\D' \equiv \lambda x\of\s.\D'_1$. By induction hypothesis,
      $\lambda x\of\s.\D'_1 \taka \lambda x\of\s.\D^\ast_1 \equiv
      \D^\ast$;
    \item if $\D \equiv \D_1\at\D_2$ and $\D$ is not a $\beta$-redex,
      then, for some $\D'_1$ and $\D'_2$, we have that
      $\D_1 \taka \D'_1$, $\D_2 \taka \D'_2$ and
      $\D' \equiv \D'_1\at\D'_2$. By induction hypothesis,
      $\D' \taka \D^\ast_1\at\D^\ast_2 \equiv \D^\ast$;
    \item if $\D \equiv (\lambda x\of\s.\D_1)\at\D_2$, then, for some
      $\D'_1$ and $\D'_2$, we have that $\D_1 \taka \D'_1$,
      $\D_2 \taka \D'_2$ and we have 2 subcases:
      \begin{itemize}
      \item $\D' \equiv (\lambda x\of\s.\D'_1)\at\D'_2$: by induction
        hypothesis,
        $\D' \taka \subst{\D^\ast_1}{x}{\D^\ast_2} \equiv \D^\ast$;
      \item $\D' \equiv \subst{\D'_1}{x}{\D'_2}$: we also have
        $\D' \taka \subst{\D^\ast_1}{x}{\D^\ast_2}$, thanks to point
        $(3)$ of Lemma \ref{lem:takaprop};
      \end{itemize}
    \item if $\D \equiv \pri \D_1$ and $\D_1$ is not a strong pair,
      then, for some $\D'_1$, we have that $\D_1 \taka \D'_1$ and
      $\D' \equiv \pri \D'_1$, therefore, by induction hypothesis,
      $\D' \taka \pri \D^\ast_1 \equiv \D^\ast$;
    \item if $\D \equiv \pri \spair{\D_1}{\D_2}$, then, for some
      $\D'_1$ and $\D'_2$, we have that $\D_1 \taka \D'_1$,
      $\D_2 \taka \D'_2$ and we have 2 subcases:
      \begin{itemize}
      \item $\D' \equiv \pri \spair{\D'_1}{\D'_2}$: by induction
        hypothesis, $\D' \taka \D^\ast_i \equiv \D^\ast$;
      \item $\D' \equiv \D'_i$: we also have, by induction hypothesis,
        $\D' \taka \D^\ast_i \equiv \D^\ast$.
      \end{itemize}
    \end{itemize}
  \end{proof}
\end{lemma}
The Church-Rosser property follows.
\begin{theorem}[Confluence]\hfill\\
  If $\D_1 \multired \D_2$ and
  $\D_1 \multired \D_3$, then there exists $\D_4$ such
  that $\D_2 \multired \D_4$ and
  $\D_3 \multired \D_4$.
  \begin{proof}
    Thanks to the first two points of Lemma \ref{lem:takaprop}, we
    know that $\multired$ is the transitive closure of
    $\taka$, therefore we can deduce the confluence property of
    $\multired$ with the usual diagram chase, as suggested
    below.
    \begin{center}
      \begin{tikzpicture}[double equal sign distance,scale=1.5]
        \foreach \id/\x/\y in { a0/0/0,b0/0/1,c0/0/2,
          a1/1/0,b1/1/1,c1/1/2, a2/2/0,b2/2/1,c2/2/2,
          a3/3/0,b3/3/1,c3/3/2}
        \node (G-\id) at (\x,-\y) {$\Delta_{\x,\y}$};
        \foreach \from/\to in {a0/b0,b0/c0,a0/a1,a1/a2,a2/a3}
        { \draw[double,-implies] (G-\from) -- (G-\to); }
        \foreach \from/\to in {
          b0/b1,b1/b2,b2/b3, c0/c1,c1/c2,c2/c3, a1/b1,b1/c1,
          a2/b2,b2/c2, a3/b3,b3/c3}
        { \draw[double, dashed,-implies](G-\from) -- (G-\to); }
      \end{tikzpicture}
    \end{center}
  \end{proof}
\end{theorem}

The next lemma says that all type derivations for $\D$ have an unique
type.
\begin{lemma}[Unicity of typing]\label{UNIQUE}\hfill\\
  If $B \vdash^\TT_\R \D : \s$, then $\s$ is unique.
  \begin{proof}
    By induction on the shape of $\D$.
  \end{proof}
\end{lemma}

The next lemma proves inversion properties on typable $\D$-terms.
\begin{lemma}[Generation]\label{GEN}\hfill
  \begin{enumerate}
  \item If $B\vdash^\TT_\R x : \s$, then $x\of\s\in B$;
  \item if $B\vdash^\TT_\R \lambda x\of\s.\D : \r$, then $\r\equiv\s\to\t$
    for some $\t$ and $B,x\of\s\vdash^\TT_\R \D : \t$;
  \item if $B\vdash^\TT_\R \D_1\at\D_2 : \t$, then there is $\s$ such that
    $B\vdash^\TT_\R\D_1 : \s\to\t$ and $B\vdash^\TT_\R\D_2:\s$;
  \item if $B\vdash^\TT_\R \spair{\D_1}{\D_2} : \r$, then there is $\s,\t$
    such that $\r\equiv\s\cap\t$ and $B\vdash^\TT_\R \D_1:\s$ and
    $B\vdash^\TT_\R\D_2:\t$ and $\essence{\D_1} \mathrel{\R} \essence{\D_2}$;
  \item if $B\vdash^\TT_\R\prl\D:\s$, then there is $\t$ such that
    $B\vdash^\TT_\R\D:\s \cap\t$;
  \item if $B\vdash^\TT_\R\prr\D:\t$, then there is $\s$ such that
    $B\vdash^\TT_\R\D:\s\cap\t$;
  \item if $B\vdash^\TT_\R u_\D :\s$, then $\s\equiv\om$;
  \item if $B\vdash^\TT_\R\D^\t:\r$, then $\r\equiv\t$ and there is $\s$ such
    that $\s\leq_\TT\t$ and $B\vdash^\TT_\R\D:\s$.
  \end{enumerate}
  \begin{proof}
    The typing rules are uniquely syntax-directed, therefore we can
    immediately conclude.
  \end{proof}
\end{lemma}

The next lemma says that all subterms of a typable $\D$-term are
typable too.
\begin{lemma}[Subterms typability]\label{lem:subterms}\hfill\\
  If $B \vdash^\TT_\R \D : \s$, and $\D'$ is a subterm of $\D$, then
  there exists $B'$ and $\t$ such that $B' \supseteq B$ and
  $B' \vdash^\TT_\R \D' : \t$.
  \begin{proof}
    By induction on the derivation of $B \vdash^\TT_\R \D : \s$.
  \end{proof}
\end{lemma}

As expected, the weakening and strengthening properties on contexts
are verified.
\begin{lemma}[Free-variable properties]\label{FREE}\hfill
  \begin{enumerate}
  \item If $B \vdash^\TT_\R \D : \s$, and $B' \supseteq B$, then
    $B' \vdash^\TT_\R \D : \s$;
  \item if $B \vdash^\TT_\R \D : \s$, then ${\rm FV}(\D) \subseteq Dom(B)$;
  \item if $B \vdash^\TT_\R \D : \s$, $B' \subseteq B$ and
    ${\rm FV}(\D) \subseteq Dom(B')$, then $B' \vdash^\TT_\R \D : \s$.
  \end{enumerate}
  \begin{proof}
    By induction on the derivation of $B \vdash^\TT_\R \D : \s$.
  \end{proof}
\end{lemma}

The next lemma also says that essence is closed under substitution.
\begin{lemma}[Substitution] \hfill \label{SUB}
  \begin{enumerate}
  \item
    $\essence{\subst{\D_1}{x}{\D_2}} \equiv
    \subst{\essence{\D_1}}{x}{\essence{\D_2}}$; \label{SUBES}
  \item If $B,x\of\s \vdash^\TT_\R \D_1 : \t$ and $B \vdash^\TT_\R \D_2 : \s$, then
    $B \vdash^\TT_\R \subst{\D_1}{x}{\D_2} :\t$.
  \end{enumerate}
  \begin{proof}\hfill
    \begin{enumerate}
    \item by induction on the shape of $\D_1$;
    \item by induction on the derivation. As an illustration, we show
      the case when the last applied rule is $(\cap I)$. Then we have
      that $B,x\of\s \vdash^\TT_\R \spair{\D_1}{\D_1'} : \t \cap \t'$ and
      $B \vdash^\TT_\R \D_2 : \s$; by induction hypothesis we have
      $B \vdash^\TT_\R \subst{\D_1}{x}{\D_2} :\t$ and
      $B \vdash^\TT_\R \subst{\D_1'}{x}{\D_2} :\t'$. Moreover, thanks to
      point $(1)$, we can show that
      $\mathop{\essence{\subst{\D_1}{x}{\D_2}}} \mathrel{\R}
      \mathop{\essence{\subst{\D_1'}{x}{\D_2}}}$. As a consequence:
      \[
        \infer[(\cap I)]{B \vdash^\TT_\R \subst{\spair{\D_1}{\D_1'}}{x}{\D_2}
          : \t \cap \t'}{ B \vdash^\TT_\R \subst{\D_1}{x}{\D_2} : \t & B
          \vdash^\TT_\R \subst{\D_1'}{x}{\D_2} : \t' &
          \mathop{\essence{\subst{\D_1}{x}{\D_2}}} \mathrel{\R}
          \mathop{\essence{\subst{\D_1'}{x}{\D_2}}} }
      \]
    \end{enumerate}
  \end{proof}
\end{lemma}

In order to prove subject reduction, we need to prove that reducing
$\D$-terms preserve the side-condition
$\essence{\D_1} \mathrel{\R} \essence{\D_2}$ when typing the strong
pair $\spair{\D_1}{\D_2}$. We prove this in the following lemma.
\begin{lemma}[Essence reduction]\label{ESSRED}\hfill
  \begin{enumerate}
  \item If $B \vdash^\TT_\equiv \D_1 : \s$ and
    $\D_1 \red \D_2$, then
    $\essence{\D_1} =_{\beta} \essence{\D_2}$;
  \item for $\R \in \{=_\beta,=_{\beta\eta}\}$, if
    $B \vdash^\TT_\R \D_1 : \s$ and $\D_1 \red \D_2$, then
    $\essence{\D_1} \mathrel{\R} \essence{\D_2}$;
  \item if $B \vdash^\TT_{=_{\beta\eta}} \D_1 : \s$ and
    $\D_1 \red_\eta \D_2$, then
    $\essence{\D_1} =_{\eta} \essence{\D_2}$.
  \end{enumerate}
  \begin{proof}
    If $\D_1$ is a redex, then we have three cases:
    \begin{itemize}
    \item if $\D_1 \equiv (\lambda x \of \s. \D'_1) \at \D''_1$ and
      $\D_2$ is $\subst{\D'_1}{x}{\D''_1}$, then, thanks to Lemma
      \ref{SUB}(\ref{SUBES}) we have that
      $\essence{\D_2} \equiv
      \subst{\essence{\D'_1}}{x}{\essence{\D''_1}}$, therefore
      $\essence{\D_1} =_\beta \essence{\D_2}$;
    \item if $\D_1 \equiv \pri \spair{\D'_1}{\D'_2}$ and $\D_2$ is
      $\D'_i$, we know that $\D_1$ is typable in $\DD{\TT}{\R}$, and
      thanks to Lemma \ref{GEN}(4), we have that
      $\essence{\D'_1} \mathrel{\R} \essence{\D'_2}$. As a
      consequence, $\essence{\D_1} \mathrel{\R} \essence{\D_2}$;
    \item if $\D_1 \equiv \lambda x\of\s. \D' \at x$ with
      $x\not\in{\rm FV}(\D')$, and $\D_2$ is $\D'$, then
      $\essence{\D_1} =_\eta \essence{\D_2}$.
    \end{itemize}
    For the contextual closure, we have that
    $\D_1 \equiv \subst{\D}{x}{\D'}$, where $\D[\_]$ is a surrounding
    context and $\D'$ is a redex, and $\D_2$ is $\subst{\D}{x}{\D''}$
    where $\D''$ is the contractum of $\D'$.  Then, by Lemma
    \ref{lem:subterms} we know that $\D'$ is typable and then we
    conclude by Lemma \ref{SUB}.
  \end{proof}
\end{lemma}
The next theorem states that all the $\DD{\TT}{\equiv}$ typed systems
preserve synchronous $\beta\pri$-reduction, and all the
$\DD{\TT}{=_{\beta}}$ and $\DD{\TT}{=_{\beta\eta}}$ typed systems
preserve $\beta\pri$-reduction.
\begin{theorem}[Subject reduction for $\beta\pri$]\hfill\label{SR}
  \begin{enumerate}
  \item If $B \vdash^\TT_\equiv \D_1 : \s$ and
    $\D_1 \red^\parallel \D_2$, then
    $B \vdash^\TT_\equiv \D_2 : \s$;
  \item for $\R \in \{=_\beta,=_{\beta\eta}\}$, if
    $B \vdash^\TT_\R \D_1 : \s$ and $\D_1 \red \D_2$, then
    $B \vdash^\TT_\R \D_2 : \s$.
  \end{enumerate}
  \begin{proof}
    If $\D_1$ is a $\beta\pri$-redex, then we proceed as usual using
    Lemmas \ref{GEN} and \ref{SUB}. For the contextual closure, we
    proceed by induction on the derivation: we illustrate the most
    important case, namely $(\cap I)$ where we have to check that the
    essence condition is preserved. According to $\R$ we distinguish
    two cases:
    \begin{enumerate}
    \item (Case where $\R$ is $\equiv$). If
      $B \vdash^\TT_\equiv \spair{\D_1}{\D_2} : \s\cap\t$ and
      $\spair{\D_1}{\D_2} \red^\parallel
      \spair{\D'_1}{\D'_2}$, then
      $\essence{\D'_1} \equiv \essence{\D'_2}$ and, by induction
      hypothesis, $B \vdash^\TT_\equiv \D'_1 : \s$ and
      $B \vdash^\TT_\equiv \D'_2 : \t$, therefore
      $B \vdash^\TT_\equiv \spair{\D'_1}{\D'_2} : \s\cap\t$;
    \item (Case where $\R \in \{=_\beta,=_{\beta\eta}\}$). If
      $B \vdash^\TT_\R \spair{\D_1}{\D_2} : \s\cap\t$ and
      $\spair{\D_1}{\D_2} \red \spair{\D'_1}{\D'_2}$,
      then:
      \begin{itemize}
      \item $\essence{\D_1} \mathrel{\R} \essence{\D_2}$;
      \item by Lemma \ref{ESSRED} we have that
        $\essence{\D'_1} \mathrel{\R} \essence{\D_1}$ and
        $\essence{\D_2} \mathrel{\R} \essence{\D'_2}$;
      \item by induction hypothesis we have that
        $B \vdash^\TT_\R \D'_1 : \s$ and $B \vdash^\TT_\R \D'_2 : \t$;
      \end{itemize}
      therefore $\essence{\D'_1} \mathrel{\R} \essence{\D'_2}$ and
      $B \vdash^\TT_\R \spair{\D'_1}{\D'_2} : \s\cap\t$.
    \end{enumerate}
  \end{proof}
\end{theorem}

The next theorem states that some of the typed systems on the back of
the $\D$-chair preserve $\eta$-reduction.
\begin{theorem}[Subject reduction for $\eta$ for
  $\CDV,\BCD$]\hfill\\\label{SRETA}
  Let $\TT \in \{\CDS,\BCD\}$. If
  $B \vdash^\TT_{=_{\beta\eta}} \D_1 : \s$ and $\D_1 \red_\eta \D_2$,
  then $B \vdash^\TT_{=_{\beta\eta}} \D_2 :\s$.
  \begin{proof}
    If $\D_1$ is a $\eta$-redex, then we proceed as usual using Lemmas
    \ref{GEN} and \ref{FREE}. For the contextual closure the proof
    proceeds exactly as in Theorem \ref{SR}.
  \end{proof}
\end{theorem}

\begin{remark}[About subject expansion]\hfill\\
  We know that some of the intersection type assignment systems \ala\
  Curry (\viz\ $\LBCD$ and $\LCDS$) satisfy the subject
  $\beta$-expansion property: one may ask whether this property can
  also be meaningful in typed systems \ala\ Church. It is not
  surprising to see that the answer is negative because
  type-decorations of bound variables are hard-coded in the
  $\lambda$-abstraction and cannot be forgotten.  As a trivial example
  of the failure of the subject-expansion in all the typed systems,
  consider the following reduction:
  \[(\lambda x\of\s.x)\at(\lambda x\of\s.x) \red
    (\lambda x\of\s.x)\]
  Obviously we can type
  $\vdash^\TT_\R (\lambda x\of\s.x) : \s\to\s$ but
  $\not\,\vdash ^\TT_\R (\lambda x\of\s.x)\at (\lambda x\of\s.x) :
  \s\to\s$.
\end{remark}

\subsection{Strong normalization}
The idea of the strong normalization proof is to embed typable terms
of the $\Delta$-calculus into Church-style terms of a target system,
which is the simply-typed $\lambda$-calculus with pairs, in a
structure-preserving way (and forgetting all the essence
side-conditions). The translation is sufficiently faithful so as to
preserve the number of reductions, and so strong normalization for the
$\D$-calculus follows from strong normalization for simply-typed
$\lambda$-calculus with pairs.
A similar technique has been used in \cite{LF} to prove the strong normalization property of LF 
and in \cite{bucciarelli} to prove the strong normalization property of a subset of $\LCD$.

The target system has one atomic type called $\circ$, a special
constant term $u_\circ$ of type $\circ$ and an infinite number of
constants $c_{\s}$ of type $\s$ for any type of the target
system. We denote by $B \vdash_\times M : \s$ a typing judgment in
the target system.

\begin{definition}[Forgetful mapping]\hfill
  \begin{itemize}
  \item On intersection types.
    \[
      \begin{array}{rcl}
        \map{{\tt a}_i} & \eqdef & \circ \quad
                                   \forall {\tt a}_i \in\Adef
                                   \phantom{\hspace{2.1cm}}\\[2mm]
        \map{\s \cap \t} &\eqdef &\map{\s} \times \map{\t}\\[2mm]
        \map{\s \to \t} & \eqdef & \map{\s} \to \map{\t}
      \end{array}
    \]
    The map can be easily extended to basis $B$.
  \item On $\D$-terms.
    \begin{displaymath}
      \begin{array}{rcl}
        \map{x}_B & \eqdef & x\\[2mm]
        \map{u_\D}_B & \eqdef & u_\circ \\[2mm]
        \map{\lambda x\of\s.\D}_B & \eqdef
                           &
                             \lambda x. \map{\D}_{B,x\of\s} \\[2mm]
        \map{\D_1\at\D_2}_B & \eqdef
                           & \map{\D_1}_B \at \map{\D_2}_B \\[2mm]
        \map{\spair{\D_1}{\D_2}}_B & \eqdef
                           & \pair{\map{\D_1}_B}{\map{\D_2}_B} \\[2mm]
        \map{\pri \D}_B & \eqdef & \pri \map{\D}_B \\[2mm]
        \map{\D^\t}_B & \eqdef & c_{\map{\s}\to\map{\t}} \at \map{\D}_B
                                 \quad\mbox{if $B\vdash^\TT_\R \D:\s$}
      \end{array}
    \end{displaymath}
  \end{itemize}
\end{definition}

The following technical lemma states some properties of the forgetful
function.
\begin{lemma}\hfill\label{TECH}
  \begin{enumerate}
  \item If $B\vdash^\TT_\R\D:\s$, then $\map{\D}_B$ is defined, and,
    for all $B' \supseteq B$, $\map{\D}_B \equiv \map{\D}_{B'}$;
  \item $\map{\D_1[\D_2/x]}_B \equiv \map{\D_1}_B[\map{\D_2}_B/x]$;
  \item If $\D_1 \red \D_2$, then
    $\map{\D_1}_B \red \map{\D_2}_B$;
  \item If $B \vdash^\TT_\R \D : \s$ then
    $\map{B} \vdash_\times \map{\D}_B : \map{\s}$.
  \end{enumerate}
  \begin{proof}\hfill
    \begin{enumerate}
    \item by induction on the derivation;
    \item by induction on $\D_1$. The only interesting part is
      $\D_1 \equiv \lambda y\of\s. \D'_1$: by induction hypothesis, we
      have that
      $\map{\D'_1[\D_2/x]}_{B,x\of\s} \equiv
      \map{\D'_1}_{B,x\of\s}[\map{\D_2}_{B,x\of\s}/x]$. Therefore, we
      see that
      $\map{(\lambda y\of\s. \D'_1)[\D_2/x]}_B \equiv \lambda
      y\of\s.\map{\D'_1[\D_2/x]}_{B,x\of\s} \equiv \lambda
      y\of\s. \map{\D'_1}_{B,x\of\s}[\map{\D_2}_{B,x\of\s}/x]$, but,
      from point (1), we know that
      $\map{\D_2}_{B,x\of\s} \equiv \map{\D_2}_{B}$, and we conclude;
    \item by induction on the context of the redex;
    \item by induction on the derivation.
    \end{enumerate}
  \end{proof}
\end{lemma}

Strong normalization follows easily from the above lemmas.
\begin{theorem}[Strong normalization]\label{SN} \hfill\\
  If $B \vdash^\TT_\R \D : \s$, then $\D$ is strongly normalizing.
  \begin{proof}
    Using Lemma \ref{TECH} and the strong normalization of the simply
    typed $\lambda$-calculus with cartesian pairs.
  \end{proof}
\end{theorem}



\section{Typed systems \ala\ Church \vs\ type assignment systems \ala\
  Curry}\label{meta-2}
\subsection{Relation between type assignment systems $\LTT$ and typed
  systems $\DD{\TT}{\R}$}
It is interesting to state some relations between type assignment
systems \ala\ Church and typed systems \ala\ Curry. An interesting
property is the one of isomorphism, namely the fact that whenever we
assign a type $\s$ to a pure $\lambda$-term $M$, the same type can be
assigned to a $\D$-term such that the essence of $\D$ is
$M$. Conversely, for every assignment of $\s$ to a $\D$-term, a valid
type assignment judgment of the same type for the essence of $\D$ can
be derived.

Soundness, completeness and isomorphism between intersection typed
systems for the $\D$-calculus and the corresponding intersection type
assignment systems for the $\lambda$-calculus are defined as follows.
\begin{definition}[Soundness, completeness and
  isomorphism]\hfill\\\label{SCI}
  Let $\DD{\TT}{\R}$ and $\LTT$.
  \begin{enumerate}
    \itemsep 2mm
  \item \label{SOUND}(Soundness, $\DD{\TT}{\R} \triangleleft
    \LTT$). $B \vdash^\TT_\R \D : \s$ implies
    $B \vdash^\TT_\cap \essence{\D} : \s$;
  \item (Completeness, $\DD{\TT}{\R} \triangleright \LTT$).
    $B \vdash^\TT_\cap M :\s $ implies there exists $\D$ such that
    $M\equiv\essence{\D}$ and $B \vdash^\TT_\R \D : \s$;
  \item (Isomorphism, $\DD{\TT}{\R} \sim \LTT$).
    $\DD{\TT}{\R} \triangleright \LTT$ and
    $\DD{\TT}{\R} \triangleleft \LTT$.
  \end{enumerate}
\end{definition}

The following properties and relations between typed and type
assignment systems can be verified.
\begin{theorem}[Soundness, completeness and isomorphism]\label{ISO}\hfill\\
  The following properties between $\D$-calculi and type assignment
  systems $\LTT$ are verified.
  \[
    \begin{array}{|l|c|c|}
      \hline\\[-4.5mm]
      \DD{\TT}{\R} & \DD{\TT}{\R} \triangleleft \LTT
      & \DD{\TT}{\R} \triangleright \LTT
      \\[.5mm]\hline
      \DD{CD}{\equiv} & \surd & \surd
      \\[.5mm]
      \DD{CDV}{\equiv} & \surd & \surd
      \\[.5mm]
      \DD{CDS}{\equiv} & \surd & \surd
      \\[.5mm]
      \DD{BCD}{\equiv} & \surd & \surd
      \\[.5mm]\hline
      \DD{CD}{=_\beta} & \times & \surd
      \\[.5mm]
      \DD{CDV}{=_\beta} & \times & \surd
      \\[.5mm]
      \DD{CDS}{=_\beta} & \surd & \surd
      \\[.5mm]
      \DD{BCD}{=_\beta} & \surd & \surd
      \\[.5mm]\hline
      \DD{CDV}{=_{\beta\eta}} & \times & \surd
      \\[.5mm]
      \DD{BCD}{=_{\beta\eta}} & \times & \surd
      \\[.5mm]\hline
    \end{array}
  \]

  \begin{proof}\hfill
    \begin{enumerate}
    \item[($\triangleleft$)]
      Soundness for $\DD{\TT}{\equiv}$. Let $\D$ be such that $B \vdash^\TT_\equiv \D :\s$. We proceed by induction on the derivation. All cases proceed straightforwardly since all rules of the type and subtype system $\vdash^\TT_\equiv$ correspond exactly to the rules of the same name in the corresponding type assignment system $\vdash^\TT_\cap$ and in the same type theory $\TT$. Therefore $M \equiv \essence{\D}$ can be easily be defined and derived with type $\s$.\\[1mm]
      Soundness for $\DD{\{CDS,BCD\}}{=_\beta}$. Let $\TT \in \{\CDS,\BCD\}$. We know, thanks to \cite{bar2013} (Figure 14.2), that the following rule is admissible for $\LTT$:
      \[
        \infer[(\cap I)_{\rm adm}] {B\vdash^\TT_\cap M : \s \cap \t}
        {B \vdash^\TT_\cap M : \s & B \vdash^\TT_\cap N : \t & M
          =_\beta N}
      \]
      Then the proof
      proceeds by induction on the derivation of
      $B \vdash^\TT_{=_\beta} \D :\s$. The most important case is when
      the last used rule is $(\cap I)$: by induction we get
      $B \vdash^\TT_\cap \essence{\D_1} : \s$, and
      $B \vdash^\TT_\cap \essence{\D_2} : \t$, and
      $\essence{\D_1} =_\beta \essence{\D_2}$, and, by the essence
      definition,
      $\essence{\spair{\D_1}{\D_2}} =_\beta \essence{\D_1}$. Apply
      rule $(\cap I)_{\rm adm}$ and conclude with
      $B \vdash^\TT_\cap \essence{\D_1} : \s \cap \t$.

   \item[($\not\!\triangleleft$)]
   Loss of soundness for $\DD{CD}{=_\beta}$ and $\DD{CDV}{=_\beta}$.
   Let $\TT \in \{ \CD, \CDV\}$.
   Let ${\sf S}  \eqdef \lambda x.\lambda y.\lambda z. x \at z \at (y \at z)$ and ${\sf K} \eqdef \lambda x.\lambda y. x$.
      Let $\Delta \eqdef (\lambda x \of (\s\to\t\to\r)\to((\s\to\t)\to\s\to\r)\to\s\to\t\to\r.\lambda y\of (\s\to\t\to\r)\to(\s\to\t)\to\s\to\r.\lambda z\of \s\to\t\to\r. x\at z\at (y \at z)) (\lambda x\of \s\to\t\to\r.\lambda y\of (\s\to\t)\to\s\to\r. x) (\lambda x\of\s\to\t\to\r.\lambda y\of\s\to\t. \lambda z\of\s. x\at z\at (y\at z))$. $\Delta$ is a simply-typed term of type $(\s\to\t\to\r)\to(\s\to\t\to\r)$, and its essence is $\essence{\Delta} \equiv {\sf S\at K\at S}$. Consider the following counter-example:
      \[
        \infer{
          \vdash^{\TT}_{=_\beta} \prr \spair{\Delta}{\lambda x\of\s.x} :
          \s \to \s
        }
        {
          \infer
          {
            \vdash^{\TT}_{=_\beta} \spair{\Delta}{\lambda x\of\s.x} :
            ((\s \to \t \to \r) \to (\s \to \t \to \r)) \cap (\s \to \s)
          }
          {
            \infer{\vdash^{\TT}_{=_\beta} \Delta : (\s \to \t \to \r) \to (\s \to \t \to \r)}{\vdots} &
            \infer{
              \vdash^{\TT}_{=_\beta} \lambda x\of\s.x : \s\to\s
            }{
              x\of\s\vdash^{\TT}_{=_\beta} x : \s
            }
            & {\sf S\at K\at S} =_\beta \lambda x.x
          }
        }
      \]
      The essence of $\prr \spair{\Delta}{\lambda x\of\s.x}$ is ${\sf S\at K\at S}$, but, if $\s$ is an atomic type:
      \vspace{-1mm}
      \[\not\,\vdash^\TT_{\cap}{\sf S\at K\at S} :
        \s\to\s \]

    \noindent Loss of soundness in
      $\DD{CDV}{=_{\beta\eta}}$ is proved via the following
      counterexample, where\\ $B \eqdef \{  x\of(\s\to\t)\cap\r \}$.

      \[
        \infer{B\vdash^{\CDV}_{=_{\beta\eta}}\prr\spair{\lambda
            y\of\s.  (\prl x)\at y}{\prr x} : \r} {
          \infer{B\vdash^{\CDV}_{=_{\beta\eta}} \spair{\lambda
              y\of\s.(\prl x)\at y}{\prr x} : (\s\to\t)\cap\r}{
            \infer{B\vdash^{\CDV}_{=_{\beta\eta}} \lambda y\of\s.(\prl
              x)\at y : \s\to\t}{
              \infer{B,y\of\s\vdash^{\CDV}_{=_{\beta\eta}}(\prl x)\at
                y : \t}{
                \infer{B,y\of\s\vdash^{\CDV}_{=_{\beta\eta}}\prl x :
                  \s\to\t }{
                  B,y\of\s\vdash^{\CDV}_{=_{\beta\eta}}x:(\s\to\t)\cap\r
                } & B,y\of\s\vdash^{\CDV}_{=_{\beta\eta}}y : \s } } &
            \infer{B\vdash^{\CDV}_{=_{\beta\eta}}\prr x : \r}{
              B\vdash^{\CDV}_{=_{\beta\eta}} x:(\s\to\t)\cap\r } &
            \lambda y.x\at y =_{\beta\eta} x } }
      \]
      The essence of
      $\prr\spair{\lambda y\of\s.(\prl x)\at y}{\prr x}$ is
      $\lambda y.x\at y$, but, if $\r$ is an atomic type:
      \vspace{-1mm}
      \[x\of(\s\to\t)\cap\r\not\,\vdash^\CDV_{\cap}\lambda y.x\at y :
        \r \]

      \noindent Loss of soundness in $\DD{BCD}{=_{\beta\eta}}$ is proved via the
      following counterexample:
      \begin{displaymath}
        \infer
        {x\of\s\vdash^\BCD_{=_{\beta\eta}}
          \prr \spair{\lambda y\of\om.x^{\om\to\om}\at y}{x} : \s}
        {
          \infer{x\of\s\vdash^\BCD_{=_{\beta\eta}}
            \spair{\lambda y\of\om.x^{\om\to\om}\at y}{x} : (\om\to\om) \cap \s}{
            \infer{x\of\s\vdash^\BCD_{=_{\beta\eta}}
              \lambda y\of\om.x^{\om\to\om}\at y : \om\to\om}{
              \infer{x\of\s,y\of\om\vdash^\BCD_{=_{\beta\eta}}
                x^{\om\to\om}\at y : \om}{
                \infer{x\of\s,y\of\om\vdash^\BCD_{=_{\beta\eta}}
                  x^{\om\to\om} : \om\to\om
                }{
                  x\of\s,y\of\om\vdash^\BCD_{=_{\beta\eta}}x : \s
                  & \s \leq_\TT \om\to\om
                }
                & x\of\s,y\of\om\vdash^\BCD_{=_{\beta\eta}} y : \om
              }
            }
            & x\of\s\vdash^\BCD_{=_{\beta\eta}}x : \s
            & \lambda y.x\at y =_{\beta\eta} x
          }
        }
      \end{displaymath}
      The essence of
      $\prr \spair{\lambda y\of\om.x^{\om\to\om}\at y}{x}$ is
      $\lambda y.x\at y$, but, if $\s$ is an atomic type (different
      than $\om$):
      \[x\of\s\not\,\vdash^\BCD_{\cap}\lambda y.x\at y : \s\]
    \item[($\triangleright$)] Let $M$ be such that
      $B \vdash^\TT_\cap M : \s$ for a given $B$. We proceed by
      induction on the derivation. All cases proceed straightforwardly
      since all rules of the type and subtype assignment system
      $\vdash^\TT_\cap$ correspond exactly to the rules of the same
      name in the corresponding typed system $\vdash^\TT_\R$ and in the
      same type theory $\TT$. Therefore a $\D$-term can be easily be
      constructed and derived with type $\s$;
    \end{enumerate}
  \end{proof}
\end{theorem}

The last theorem characterizes the class of strongly normalizing
$\D$-terms.
\begin{theorem}[Characterization]\hfill\\
  Every strongly normalizing $\lambda$-term can be type-annotated so
  as to be the essence of a typable $\D$-term.
  \begin{proof}
    We know that every strongly normalizing $\lambda$-term $M$ is
    typable in $\LTT$. By Theorem \ref{ISO} we have that
    $\DD{\TT}{\R} \triangleright \LTT$, therefore there exists some
    typable $\D$, such that $M \equiv \essence{\D}$.
  \end{proof}
\end{theorem}

We can finally state decidability of type checking (TC) and type
reconstruction (TR).
\begin{theorem}[Decidability of type checking and type reconstruction]\label{TCTR}\hfill\\
  \[
    \begin{array}{|l|c|}
      \hline\\[-4.5mm]
      \DD{\TT}{\R} & {\rm TC/TR}
      \\[.5mm]\hline
      \DD{CD}{\equiv}
                   & \surd
      \\[.5mm]
      \DD{CDV}{\equiv}
                   & \surd
      \\[.5mm]
      \DD{CDS}{\equiv}
                   & \surd
      \\[.5mm]
      \DD{BCD}{\equiv}
                   & \surd
      \\[.5mm]\hline
      \DD{CD}{=_\beta}
                   & \surd
      \\[.5mm]
      \DD{CDV}{=_\beta}
                   & \surd
      \\[.5mm]
      \DD{CDS}{=_\beta}
                   & \times
      \\[.5mm]
      \DD{BCD}{=_\beta}
                   & \times
      \\[.5mm]\hline
      \DD{CDV}{=_{\beta\eta}}
                   & \surd
      \\[.5mm]
      \DD{BCD}{=_{\beta\eta}}
                   & \times
      \\[.5mm]\hline
    \end{array}
  \]
  \begin{proof}
    Both type checking and type reconstruction can be proved by
    induction on the structure of $\D$, using the decidability of
    $\BCD$ proved by Hindley \cite{Hindley82} (see also
    \cite{TTCS17}). By Theorem \ref{ISO}, the essences of all the
    $\D$-terms, which are typable in $\DD{CD}{=_\beta}$,
    $\DD{CDV}{=_\beta}$, or $\DD{CDV}{=_{\beta\eta}}$, are typable in
    $\LCD$ or $\LCDV$, therefore they are strongly normalizing. As a
    consequence, the side-condition
    $\essence{\D_1} \mathrel{\R} \essence{\D_2}$ is decidable for
    $\DD{CD}{=_\beta}$, $\DD{CDV}{=_\beta}$, and
    $\DD{CDV}{=_{\beta\eta}}$ and so type reconstruction and type
    checking are decidable too.

    Type reconstruction and type checking are not decidable in
    $\DD{CDS}{=_\beta}$, $\DD{BCD}{=_\beta}$, and
    $\DD{BCD}{=_{\beta\eta}}$, because $\spair{u_{\D_1}}{u_{\D_2}}$ is
    typable if and only if $\essence{\D_1} =_\beta \essence{\D_2}$
    (resp. $\essence{\D_1} =_{\beta\eta} \essence{\D_2}$). However,
    $\essence{\D_1}$ and $\essence{\D_2}$ are arbitrary pure
    $\lambda$-terms, and both $\beta$-equality and
    $\beta\eta$-equality are undecidable.
  \end{proof}
\end{theorem}

\subsection{Subtyping and explicit coercions}\label{tannen}
The typing rule $(\leq_\TT)$ in the general typed system introduces
type coercions: once a type coercion is introduced, it cannot be
eliminated, so {\it de facto} freezing a $\D$-term inside an explicit
coercion.  Tannen \etal\ \cite{TannenCGS91} showed a translation of a
judgment derivation from a ``Source" system with subtyping (Cardelli's
Fun \cite{CardFun}) into an ``equivalent'' judgment derivation in a
``Target" system without subtyping (Girard system F with records and
recursion).  In the same spirit, we present a translation that removes
all explicit coercions.  Intuitively, the translation proceeds as
follows: every derivation ending with rule
\[
  \infer[(\leq_\TT)] {B \vdash^\TT_\R \D^\t : \t} {B \vdash^\TT_\R \D
    : \s & \s \leq_\TT \t}
\]
is translated into the following (coercion-free) derivation
\[
  \infer[({\to}E)] {B \vdash^\TT_{\R'} \trans{\s \leq_\TT \t} \at
    \trans{\D}_B : \t} {B \vdash^\TT_{\R'} \trans{\s \leq_\TT \t} : \s
    \to \t & B \vdash^\TT_{\R'} \trans{\D}_B : \s}
\]
where $\R'$ is a suitable relation such that $\R \sqsubseteq
\R'$. Note that changing of the type theory is necessary to guarantee
well-typedness in the translation of strong pairs. Summarizing, we
provide a type preserving translation of a $\D$-term into a
coercion-free $\D$-term such that
$\essence{\D} =_{\beta\eta} \essence{\D'}$.


The following example illustrates some trivial compilations of axioms and rule schemes of Figure \ref{subtype}.
\begin{example}[Translation of axioms and rule schemes of Figure \ref{subtype}]
  \hfill
  \begin{enumerate}
  \item[(refl)] the judgment $x \of \s \vdash^\TT_\R  \spair{x}{x^\s} : \s \cap \s$ is
    translated to a coercion-free judgment
    \[ x \of \s \vdash^\TT_{=_\beta} \spair{x}{(\lambda y\of \s.y) \at x} : \s \cap \s \]
  \item[(incl)] the judgment
    $x \of \s \cap \t \vdash^\TT_\R \spair{x}{x^\t} : (\s \cap \t) \cap \t$ is
    translated to a coercion-free judgment
    \[ x \of \s \cap \t \vdash^\TT_{=_\beta} \spair{x}{(\lambda y\of
        \s \cap \t. \prr y) \at x} : (\s \cap \t) \cap \t \]
  \item[(glb)] the judgment
    $x \of \s \vdash^\TT_\R \spair{x}{x^{\s\cap\s}} : \s \cap (\s
    \cap\s)$ is translated to a coercion-free judgment
    \[ x \of \s \vdash^\TT_{=_\beta} \spair{x}{(\lambda y\of
        \s. \spair{y}{y}) \at x} : \s \cap (\s \cap \s) \]
  \item[$(\om_{top})$] the judgment
    $x\of\s \vdash^\TT_\R \spair{x}{x^\om} : \s \cap \om $ is translated to a
    coercion-free judgment
    \[ x\of\s \vdash^\TT_{=_\beta} \spair{x}{(\lambda y\of\s.u_y)\at x} : \s \cap \om \]
  \item[$(\om_{\to})$] the judgment
    $x\of\om \vdash^\TT_\R \spair{x}{x^{\s\to\om}} : \om \cap (\s\to\om)$ is translated to
    a coercion-free judgment
    \[ x\of\om \vdash^\TT_{=_{\beta\eta}} \spair{x}{(\lambda f\of\om.\lambda
      y\of\s. u_{(f\at y)})\at x} : \om \cap (\s\to\om) \]
  \item[$({\to}\cap)$] the judgment
    $x\of(\s\to\t)\cap(\s\to\r) \vdash^\TT_\R x^{\s\to\t\cap\r} :
    \s\to \t\cap\r$ is translated to a coercion-free judgment
    \[ x\of(\s\to\t)\cap(\s\to\r) \vdash^\TT_{=_{\beta\eta}} (\lambda
      f\of(\s\to\t)\cap(\s\to\r).\lambda y\of\s.\spair{(\prl f)\at
        y}{(\prr f)\at y})\at x : \s\to \t\cap\r \]
  \item[$(\to)$] the judgment
    $x\of\s \to \t\cap\r \vdash^\TT_\R \spair{x}{x^{\s\cap \r \to\t}}: (\s \to \t\cap\r) \cap (\s\cap \r
    \to\t)$ is translated to a coercion-free judgment
    \[ x\of\s\to \t\cap\r \vdash^\TT_{=_{\beta\eta}} \spair{x}{(\lambda
      f\of\s\to \t\cap\r. \lambda y\of\s\cap\r.\prl (f\at (\prl
      y)))\at x} : (\s \to \t\cap\r) \cap (\s\cap \r  \to\t) \]
  \item[(trans)] the judgment
    $x \of \s \vdash^\TT_\R \spair{x}{(x^\om)^{\s \to\om}} : \s \cap (\s \to \om)$ is
    translated to a coercion-free judgment
    \[ x \of \s \vdash^\TT_{=_{\beta\eta}} \spair{x}{(\lambda f\of \om.\lambda
      y\of \s.u_{(f \at y)}) \at ((\lambda y\of\s.u_y) \at x)} : \s \cap (\s \to \om) \]
  \end{enumerate}
\end{example}

The next definition introduces two maps translating subtype judgments
into explicit coercions functions and $\D$-terms into coercion-free
$\Delta$-terms.
\begin{definition}[Translations $\trans{-}$ and $\trans{-}_B$]\hfill
  \begin{enumerate}
  \item The minimal type theory $\leqmin$ and the extra axioms and
    schemes are translated as follows.
    \[
      \begin{array}[c]{lrcl} 
      {\rm (refl)}
        &
        \trans{\s \leq_\TT \s}
        & \eqdef
        & \vdash^\TT_{=_\beta}\lambda x\of\s.x : \s\to\s
        \\[2mm]
        {\rm (incl_1)}
        &
          \trans{\s \cap \t \leq_\TT \s}
        & \eqdef
        & \vdash^\TT_{=_\beta}\lambda x\of\s\cap\t.\prl x :
          \s \cap\t \to \s
        \\[2mm]
        {\rm (incl_2)}
        &
          \trans{\s \cap \t \leq_\TT \t}
        & \eqdef
        & \vdash^\TT_{=_\beta}\lambda x\of\s\cap\t.\prr x :
          \s \cap\t \to \t
        \\[2mm]
        {\rm (glb)}
        &
          \trans{\dfrac{\r \leq_\TT \s \quad \r \leq_\TT \t}
          {\r \leq_\TT \s \cap \t}}
        & \eqdef
        & {\vdash^\TT_{=_\beta}\lambda x\of\r.
          \spair{\trans{\r \leq_\TT \s} x}{\trans{\r \leq_\TT \t} x} :
          \r \to \s \cap \t}
        \\[4mm]
        {\rm (trans)}
        &
          \trans{\dfrac{\s \leq_\TT \t \quad \t \leq_\TT \r}{\s \leq_\TT \r}}
        & \eqdef
        & {\vdash^\TT_{=_\beta} \lambda x\of\s. \trans{\t \leq_\TT \r}
          \at (\trans{\s \leq_\TT \t} \at x) : \s \to \r}
        \\[4mm]
        (\om_{top})
        &
          \trans{\s \leq_\TT\om}
        &\eqdef
        & \vdash^\TT_{=_\beta}\lambda x\of\s.u_x : \s\to\om
        \\[2mm]
        (\om_\to)
        &
          \trans{\om \leq_\TT \s \to \om}
        & \eqdef
        & \vdash^\TT_{=_{\beta\eta}}\lambda f\of\om.\lambda x\of \s.
          u_{(f\at x)} : \om\to(\s\to\om)
        \\[2mm]
        \multicolumn{4}{l}{\rm Let~\xi_1 \eqdef (\s \to \t) \cap (\s \to \r)
        ~and~\xi_2 \eqdef \s \to \t \cap \r}
        \\[2mm]
        ({\to}\cap)
        &
          \trans{\xi_1 \leq_\TT \xi_2}
        &\eqdef
        & \vdash^\TT_{=_{\beta\eta}} \lambda f\of\xi_1.\lambda x\of\s.
          \spair{(\prl f)\at x}{(\prr f)\at x} : \xi_1 \to \xi_2
        \\[4mm]
        \multicolumn{4}{l}{\rm Let~\xi_1 \eqdef \s_1 \to \t_1
        ~and~\xi_2 \eqdef \s_2 \to \t_2}
        \\[2mm]
        (\to)
        &
          \trans{\dfrac{\s_2\leq_\TT \s_1 \quad \t_1 \leq_\TT\t_2}
          {\s_1 \to\t_1\leq_\TT\s_2 \to\t_2}}
        & \eqdef
        & \vdash^\TT_{=_{\beta\eta}} \lambda f\of\xi_1.\lambda x\of\s_2.
          \trans{\t_1\leq_\TT \t_2}
         \at  (f\at(\trans{\s_2\leq_\TT \s_1} x)) : \xi_1 \to \xi_2
        \\[4mm]
      \end{array}
    \]
  \item The translation $\trans{-}_B$ is defined on $\D$ as follows.
    \begin{eqnarray*}
      \trans{u_\D}_B
      & \eqdef
      & u_{\trans{\D}_B}
      \\[2mm]
      \trans{x}_B
      & \eqdef & x
      \\[2mm]
      \trans{\lambda x\of\s.\D}_B
      & \eqdef
      & \lambda x\of\s.\trans{\D}_{B,x\of \s}
      \\[2mm]
      \trans{\D_1 \at \D_2}_B
      & \eqdef
      & \trans{\D_1}_B \at \trans{\D_2}_B
      \\[2mm]
      \trans{\spair{\D_1}{\D_2}}_B
      & \eqdef
      & \spair{\trans{\D_1}_B}{\trans{\D_2}_B}
      \\[2mm]
      \trans{\pri \D}_B
      & \eqdef
      & \pri \trans{\D}_B \qquad\qquad \,\, i \in \{1,2\}
      \\[2mm]
      \trans{\D^\t}_B
      & \eqdef
      & \trans{\s \leq_\TT \t} \at \trans{\D}_B \quad
        \IF B\vdash^\TT_\R \D :\s.
    \end{eqnarray*}
  \end{enumerate}
\end{definition}

By looking at the above translation functions we can see that if
$B \vdash^\TT_\R \D : \s$, then $\trans{\D}_B$ is defined and it is
coercion-free.

The following lemma states that a coercion function is always typable
in $\DD{\TT}{=_{\beta\eta}}$, that it is essentially the identity and
that, without using the rule schemes $({\to}\cap)$, $(\om_{\to})$, and $(\to)$
the translation can even be derivable in $\DD{\TT}{=_\beta}$.

\begin{lemma}[Essence of a coercion is an identity]\hfill\label{coerce}
  \begin{enumerate}
  \item If $\s \leq_\TT \t$, then
    $\vdash^\TT_{=_{\beta\eta}} \trans{\s\leq_\TT \t} : \s \to \t$ and
    $\essence{\trans{\s\leq_\TT \t}} =_{\beta\eta} \lambda x.x$;
  \item If $\s \leq_\TT \t$ without using the rule schemes
    $({\to}\cap)$, $(\om_{\to})$, and $(\to)$, then
    $\vdash^\TT_{=_\beta} \trans{\s\leq_\TT \t} : \s \to \t$ and
    $\essence{\trans{\s\leq_\TT \t}} =_\beta \lambda x.x$.
  \end{enumerate}
  \begin{proof}
    The proofs proceed in both parts by induction on the derivation of
    $\s \leq_\TT \t$. For instance, in  case of  (glb), we can verify
    that
    $\vdash^\TT_{=_\beta}\lambda x\of\r.\spair{\trans{\r \leq_\TT \s}
      x}{\trans{\r \leq_\TT \t} x} : \r \to \s \cap \t$ using the
    induction hypotheses that $\trans{\r \leq_\TT \s}$ (resp.
    $\trans{\r \leq_\TT \t}$) has type $\r\to\s$ (resp. $\r\to\t$) and
    has an essence convertible to $\lambda x.x$.
  \end{proof}
\end{lemma}

We can now prove the coherence of the translation as follows.
\begin{theorem}[Coherence]\hfill\\
  If $B \vdash^\TT_\R \D : \s$, then
  $B \vdash^\TT_{\R'} \trans{\D}_B : \s$ and
  $\essence{\trans{\D}_B} \mathrel{\R'} \essence{\D}$, where
  $\DD{\TT}{\R}$ and $\DD{\TT}{\R'}$ are respectively the source and
  target intersection typed systems given in Figure \ref{nosub}.
  \begin{proof}
    By induction on the derivation. We illustrate the most important
    case, namely when the last type rule is $(\leq_\TT)$. In this case
    $\trans{\D^\t}_B$ is translated to
    $\trans{\s \leq_\TT \t} \at \trans{\D}_B$. By induction hypothesis
    we have that $B \vdash^\TT_\R \D : \s$, and by Lemma \ref{coerce} we
    have that $B \vdash^\TT_{\R'} \trans{\s \leq_\TT \t} : \s\to\t$;
    therefore $B \vdash^\TT_{\R'} \trans{\D^\t}_B : \t$. Moreover, we
    know that
    $\essence{\trans{\s \leq_\TT \t}} \mathrel{\R'} \lambda x.x$, and
    this gives
    $\essence{\trans{\D^\t}_B} \mathrel{\R'}
    \essence{\trans{\D}_B}$. Again by induction hypothesis we have that
    $\essence{\trans{\D}_B} \mathrel{\R'} \essence \D$, and this gives
    the thesis
    $\essence{\trans{\D^\t}_B} \mathrel{\R'} \essence{\D^\t}$.
  \end{proof}
\end{theorem}

\begin{figure}[t]
  \[
    \begin{array}{|l|l|l|}
      \hline\\[-4.5mm]
      \mbox{Source} & \mbox{Target}
      \\[.5mm]\hline
      \DD{CD}{\equiv}
                    &
                      \DD{CD}{=_\beta}
      \\[.5mm]
      \DD{CDV}{\equiv}
                    &
                      \DD{CDV}{=_{\beta\eta}}
      \\[.5mm]
      \DD{CDS}{\equiv}
                    &
                      \DD{CDS}{=_{\beta}}
      \\[.5mm]
      \DD{BCD}{\equiv}
                    &
                      \DD{BCD}{=_{\beta\eta}}
      \\[.5mm]\hline
      \DD{CD}{=_\beta}
                    &
                      \DD{CD}{=_\beta} 
      \\[.5mm]
      \DD{CDV}{=_\beta}
                    &
                      \DD{CDV}{=_{\beta\eta}}
      \\[.5mm]
      \DD{CDS}{=_\beta}
                    &
                      \DD{CDS}{=_{\beta}} 
      \\[.5mm]
      \DD{BCD}{=_\beta}
                    &
                      \DD{BCD}{=_{\beta\eta}}
      \\[.5mm]\hline
      \DD{CDV}{=_{\beta\eta}}
                    &
                      \DD{CDV}{=_{\beta\eta}} 
      \\[.5mm]
      \DD{BCD}{=_{\beta\eta}}
                    &
                      \DD{BCD}{=_{\beta\eta}} 
      \\[.5mm]\hline
    \end{array}
  \]
  \caption{On the left: source systems. On the right: target systems
    without the $(\leq_\TT)$ rule.}
  \label{nosub}
\end{figure}


\end{document}